# Superionic UO₂: A Model Anharmonic Crystalline Material

*Hao Zhang[1] [†], Xinyi Wang[1], Alexandros Chremos[2] and Jack F. Douglas[2] [†]*

[1]Department of Chemical and Materials Engineering, University of Alberta, Edmonton, Alberta T6G 1H9, Canada

[2]Material Science and Engineering Division, National Institute of Standards and Technology, Gaithersburg, Maryland 20899, USA

**Abstract:**

Crystalline materials at elevated temperatures and pressures can exhibit properties more reminiscent of simple liquids than ideal crystalline materials. Superionic crystalline materials having a liquid-like conductivity $\sigma$ are particularly interesting for battery, fuel cell, and other energy applications, and we study UO₂ as a prototypical superionic material since this material is widely studied given its commercial importance as reactor fuel. Using molecular dynamics, we first investigate basic thermodynamic and structural properties. We then quantify structural relaxation, dynamic heterogeneity, and average ions mobility. We find that the non-Arrhenius diffusion and structural relaxation time of this prototypical superionic material can be quantitatively described in terms of a generalized activated transport model ('string model') in which the activation energy varies in direct proportion to the average string length. Our transport data can also be described equally well by an Adam-Gibbs model in which the excess entropy density of the crystalline material is estimated from specific heat and thermal expansion data, consistent with the average scale of string-like collective motion scaling inversely with the excess entropy of the crystal. Strong differences in the temperature dependence of the interfacial mobility from non-ionic materials are observed, and we suggest that this difference is due to the relatively high cohesive interaction of ionic materials. In summary, the study of superionic UO₂ provides insight into the role of cooperative motion in enhancing ion mobility in ionic materials and offers design principles for the development of new superionic materials for use in diverse energy applications.

† Corresponding authors: hao.zhang@ualberta.ca; jack.douglas@nist.gov

## I. Introduction

It is common intuition that diffusion and relaxation in crystalline materials are relatively slow in comparison to liquids because of the localization of particles to lattice positions, but there are diverse and practically important materials that defy this intuition. By cooling liquids, the rate of relaxation may then occur over geologically relevant timescales, and the rate of diffusion correspondingly becomes extremely slow, a phenomenon evidenced by many plastic polymer materials, foods, and diverse everyday objects. [1] A disordered molecular structure evidently does not ensure high mobility and there is much theoretical effort currently being devoted to understanding this dynamically arrested state of matter. [2] Correspondingly, and less well appreciated by the scientific community, there are also crystalline materials for which the rates of atomic diffusion and relaxation approach values normally observed in simple fluids at elevated temperatures. These materials are termed "superionic" when the crystalline material exhibits an ionic composition and were discovered about 200 years ago by Faraday, [3, 4] but such high mobilities are predicted to arise even in regular crystalline materials, such as iron, water, ice, and ammonia at sufficiently high pressures.[5-11] Superionic materials are currently of intense interest in the development of safer, more efficient and longer lasting battery materials, and, more generally, crystalline materials of high atomic mobility are also of great interest in understanding geophysical phenomena associated with the properties of the earth's Fe core and the $MgSiO_3$ perovskite material composing a substantial fraction the earth's lower mantle [5-7, 12, 13], the lower mantle comprising over ½ the volume of the earth and dominating its properties. [14] The reactor materials $UO_2$ and $LiO_2$ are also superionic materials at elevated temperatures and a thorough understanding of the thermodynamic and dynamic properties of crystalline $UO_2$ materials is of great importance in relation to engineering reactors having improved reactor safety and efficiency.[15, 16] Other practical applications of superionic materials are mentioned below in the course of our discussion.

Although it is natural to expect that the properties of cooled liquids should progressively approach the properties of "solid" crystalline materials and that heated crystals should approach the properties of liquid materials upon heating, it is clear that something must preempt the freezing or melting transitions in these materials to allow a continuous tuning of their properties rather than the commonly found discontinuous property changes associated with freezing and melting. In these poorly understood, and often immensely useful materials, it is generally observed experimentally that relaxation and diffusion are non-Arrhenius and molecular dynamics (MD) simulations have indicated the general prevalence of molecular transport taking the form of string like collective motion of atoms. [17-20] String-like collective motion has also been evidenced in simulations of the dynamics of freezing and melting of Ni nanoparticles, [21] and in the homogeneous melting of crystals of Lennard-Jones particles [22, 23] and in crystalline Ni. [24] Simulations of a wide range of glass-forming materials under equilibrium conditions, both polymer and metallic glass materials, have further indicated that the change in the activation free energy governing diffusion and relaxation can be quantified by the change in the average cooperative exchange events ('strings') and it is currently a question whether these structures might have the same significance for heated crystals. Annamareddy and Eapen [25-27] have recently made studies of string-like collective motion in simulations of $UO_2$, $LiO_2$ and fluorite materials, where superionic dynamics has long been observed with a view of obtaining a corresponding *quantitative description* of transport in terms of string dynamics. While these simulation results are highly suggestive, they do not support a coherent picture of relaxation and diffusion that was found before in cooled liquids. [28, 29] The current work focuses on $UO_2$ as a model superionic material for which there are excellent potentials and corresponding measurement studies of diverse thermodynamic and transport properties in view of its high practical interest, [15, 16] and we analyze this model material based on a methodology used previously to quantify the dynamics of glass-forming liquids to determine what features are common and which features are specific to superionic crystalline materials, as compared to the glassy dynamics of cooled liquids. We are motivated to study this problem also as a model material for better understanding ionic transport in amorphous superionic materials, which exhibit many phenomenological parallels with superionic crystalline materials and promising superionic Li salt materials in the development of safer and longer lasting battery materials. [30-32] Finally, we mention that superionic materials are also of great interest in connection with the development of efficient thermoelectric materials [33, 34] because such materials normally have a high electrical conductivity, while at the same time having a low thermal conductivity, exactly the properties required for a low Seebeck coefficient and an efficient thermoelectric material. We note that $UO_2$ has been observed to exhibit a large Seebeck coefficient at elevated temperatures [35], another aspect of the superionic nature of $UO_2$ at high temperatures of significance for reactor operation and efficiency. [36]

## II. Model

Molecular dynamics simulations were performed to investigate thermodynamic and dynamic properties of $UO_2$ at elevated temperature using a many-body empirical potential (CRG) proposed by Qin et al.[37] This semi-empirical potential combines a pair-wise and a many-body interaction, where the pair-wise includes a long-range electrostatic interaction and short-range interactions with a combination of Buckingham and Morse forms, and the many-body is using an Embedded Atom Method (EAM). [38] This potential provides excellent overall agreement with experimental thermal expansion and enthalpy increase with temperature and reasonable pre-melting and mechanical melting with experimental values. A perfect crystal with U ions located in face-centered cubic sublattice and O ions located in simple cubic sublattice was initiated first. The simulation cell consists of 10,368 ions with a dimension of about 5.5 nm x 5 nm x 6 nm, oriented with crystallographic directions $[\bar{1}\bar{1}2]$, $[1\bar{1}0]$ and $[111]$ in the X-, Y- and Z-directions. In the simulation of bulk properties of $UO_2$, periodic boundary conditions were applied in all directions, while in the simulation interfacial diffusion, free boundary condition was applied in the Y-direction and periodic boundary conditions were applied in the other two directions. The isobaric-isothermal ensemble (NPT) was employed in bulk simulation where the zero pressure and simulation box size were controlled by the Parrinello-Rahman method[39] and constant temperature ($T$) was maintained by the Nose-Hoover method. [40, 41] The particle-particle-particle-mesh Ewald method[42] was used to ensure convergence of the long-range electrostatic energy. The MD simulations utilize LAMMPS [43], which was developed at the Sandia National Laboratories.

The simulated $UO_2$ material was first equilibrated at $T = 1500$ K for 1 ns and then continuously heated to $T = 3200$ K with a heat rate of $2 \times 10^{11}$ K / s. The potential energy and simulation box size were recorded in order to extract information on specific heat capacity and lattice parameter. Isothermal heating for an extended period of time was also applied in current study to ensure the system to reach near equilibrium and to allow us to probe kinetic processes that cannot be observed under continuous heating conditions. In current study, isothermal heating simulations were also performed at $T$ = 2400 K, 2500 K, 2600 K, 2700 K, 2800 K, 2900 K, 3000 K, 3100 K, and 3200 K. At each $T$, the simulation was conducted for at least 5 ns and up to 20 ns. Self-diffusion coefficients of O ions are determined from the slope of the mean-square-displacement versus time $t$ after long times, i.e., $D$ is defined by the long time limit of the ratio, $(\sum_{i=1}^{N}(\Delta x_i)^2 + (\Delta y_i)^2 + (\Delta z_i)^2) / 6N t$, where $N$ is the total number of O ions in the system.

## III.    Results and Discussion

### A. Basic Thermodynamic and Structural Properties of $UO_2$

Since function often follows form, it is natural to begin our discussion of our model $UO_2$ material with the temperature dependence of the pair correlation function $g(r)$ and static structure factor $S(q)$. In Figure 1, we show the evolution in the shape of $g(r)$ for the O ions of crystalline $UO_2$ as temperature is raised from $T = 2400$ K to 3000 K. The first and second peak heights drop in magnitude appreciably, and we see a splitting of the second peak below $T = 2700$ K, evidence of structural disorder, as observed also characteristically with the emergence of the glass state in liquids upon cooling. At still lower temperatures, the peaks become progressively more sharply defined and the splitting of the second peak of $g(r)$ for the O ions must eventually disappear as a crystalline order becomes more perfect. We then have a clue that these materials are exhibiting some sort of structural disordering well below the estimated equilibrium melting temperature of our simulated $UO_2$ material based on the CRG potential, $T_m = 3050$ K, determined based on the observation of point of two-phase coexistence.[37, 44] Experimentally, the equilibrium melting temperature has been estimated to equal $T_m = (3120 \pm 20$ K$)$ [45], which is reasonably consistent with the simulation estimate of the CRG model.

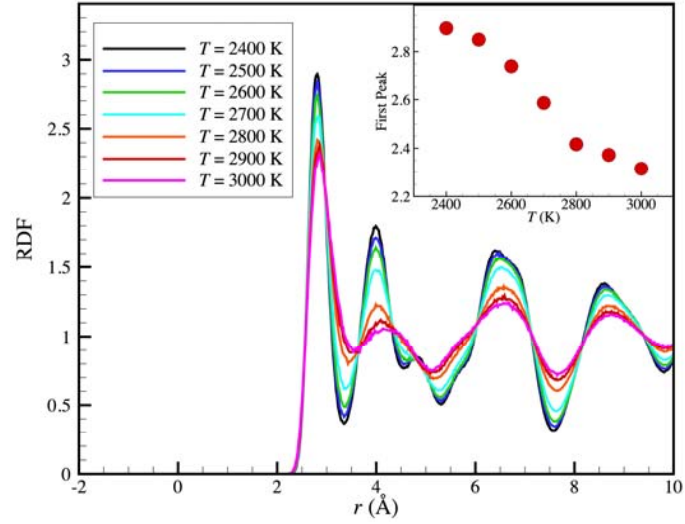

**Figure 1.** Pair correlation function $g(r)$ of O ions at different temperatures. The inset shows the first peak height of $g(r)$ as a function of temperature.

It is commonly observed that many crystalline materials begin to acquire significant molecular mobility above a definite onset temperature, [46-48] termed the 'Tammann temperature' $T_{Tam}$, and this temperature often arises in the context of practical material design. The simple rule of thumb[46-48] is that $T_{Tam}$ is normally about a factor of 2/3 times the equilibrium $T_m$, which would lead us to expect a "premelting" onset temperature for accelerated dynamics in $UO_2$ near $T = 2080$ K. Such a crossover temperature for accelerated diffusion and increase in atomic disorder in the O sub-lattice can be evidenced by examining long range order (LRO), relating directly to x-ray scattering measurements, as discussed previously by Annamareddy and Eapen[26]. LRO is quantified [49] based on a consideration of the Fourier transform of local density, i.e., $\rho(\boldsymbol{k}) = \frac{1}{N}\sum_{j=1}^{N} e^{-i\boldsymbol{k}\cdot\boldsymbol{r}_j}$, where local density $\rho(\boldsymbol{r}) = \sum_{j=1}^{N} \delta(\boldsymbol{r} - \boldsymbol{r}_j)$ , $N$ is the total number of atoms, and $\boldsymbol{k}$ is the wave vector that is chosen to be a reciprocal lattice vector of the ordered state. Based on this metric, the LRO equals unity for a perfect lattice and zero for a perfectly disordered fluid. Figure 2 shows that the degree of LRO remains stable after system equilibration and the average value of the LRO parameter at first decreases linearly with temperature, and then this quantity starts to decrease more abruptly below a temperature around 2000 K. A close examination of the lattice parameter $a(\text{Å})$, a description of the interatomic O ion spacing, shown in Figure 3, also shows a small kink around 2000 K (see inset of Figure 3). Eapen and others have identified this type of crossover in the same temperature range for $UO_2$, [26, 50] and this characteristic onset temperature has been identified in other superionics, where it is often designated as $T_\alpha$. Eapen and Annamareddy also previously noted that the onset temperature $T_\alpha$ can also be identified by a minimum in the product of $T$ times the specific heat $C_p$ divided by $T$ as a function of $T$. [51] Measurements show that Arrhenius plots of the conductivity of superionic materials exhibit a kink close to the onset temperature $T_\alpha$ [27] and a large increase in the magnitude of the self-diffusion coefficient $D_s$ of O ions at this temperature[52] so that this temperature also signals a change of the material dynamics from that of a 'simple' crystal to some kind of highly defective crystalline structure. The nature of this change of dynamics is the focus of our discussion below.

At still lower temperatures than we investigate, there is another transitional phenomenon observed in many ionic materials, including $UO_2$. In particular, there is a so-called 'intrinsic-extrinsic' transition at which there is an abrupt transition in electrical conductivity and other transport properties and Ruello et al. [4, 35] report this temperature to be 1273 K for their samples of $UO_2$. The occurrence of this transition is sensitive to sample preparation and environmental conditions, accounting for the "extrinsic" designation. [4] An interesting feature of this transition is that the activation energy increases appreciably upon passing through this transition upon heating. Below, we will show that the activation energy changes in the *opposite direction* upon passing through the superionic transition, where it decreases appreciably upon heating. Clearly, this means that the activation energy changes non-monotonically when considered over a large $T$ range. There is no general agreement on the physical origin of the intrinsic-extrinsic transition, and below we strictly confine our simulations to the "intrinsic temperature range' above 1273 K where we can readily perform our simulations. In future work, we plan to investigate the extrinsic-intrinsic transition to determine how it fits in with the superionic transition discussed below to determine if this transition also signals a change in the character of collective motion in the material.

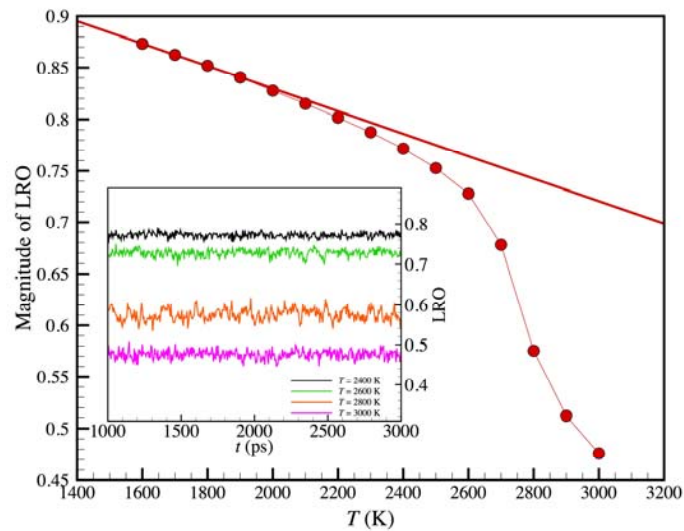

**Figure 2.** Magnitude of LRO as a function of temperature for O ions sublattice. The inset shows the temporal evolution of LRO for O ions sublattice at different temperatures with wave vector $\boldsymbol{k} = (2\pi/a)$ (010).

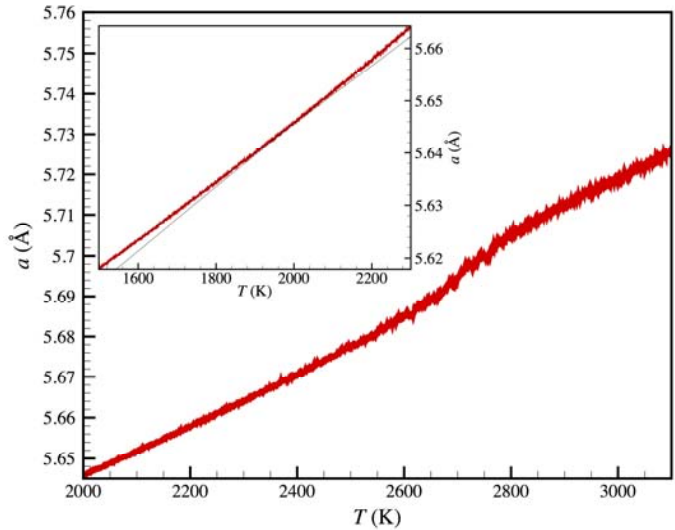

**Figure 3.** The sublattice parameter of O ions as a function of temperature. The inset shows a small kink is observed around 2000 K that defines an onset temperature $T_\alpha$ for heterogeneous structure and dynamics.

The pair correlation function $g(r)$ describes the probability that another particle being around a reference particle so that a large value of the first peak indicates strong degree of local packing. We may obtain a good sense of molecular correlations at large scales by considering the Fourier transform of $g(r)$, the (partial) static structure factor $S(q)$, and we illustrate this quantity in Figure 4. We see that $S(q)$ of the O ions develops numerous sharp peaks at low $q$, corresponding to large scales of the material. Peaks at intermediate $q$ show a substantial broadening upon cooling towards the onset temperature and the data also becomes noisier at lower temperatures. We then have further evidence of some form of mesoscale structural disorganization of our heated crystal above 2000 K. It is hard to give a clear structural interpretation to these observations, however.

Superionic crystalline materials are further conventionally classified in terms of an additional characteristic temperature that arises between $T_\alpha$ and $T_m$. Type I superionics, such Ag halide-chalcogenides (e.g., AgI, $AgS_2$, AgSI, $RbAg_4I_5$,…) and Cu halides (CuI, CuBr, CuCl), exhibit a first order solid-solid structural transition to a crystal morphology conducive to superionic transport [53-55], while type II superionics exhibit a constant pressure specific heat $C_p$ peak at a characteristic temperature designated $T_\lambda$, which has interpreted as a second order or some higher order ("rounded") thermodynamic transition. [56] This type of rounding might be rationalized to arise from the result of structural fluctuations that can convert a first order phase transition to become a continuous transition. [57] In UO$_2$, this transition is often termed the "Bredig transition" [58, 59]. "Faraday transition" [25] or lambda transition, $T_\lambda$. In Figure 5, we show $C_p$ for our model as a function of $T$ where this peak is clearly exhibited around 2700 K. For comparison, we note that Annamareddy and Eapen [26] estimated a comparable value of the lambda transition $T_\lambda$ for their potential, $T_\lambda = 2560$ K. Both computational estimates of $T_\lambda$ are reasonably consistent with the experimental estimate [58], $T_\lambda = 2610$ K. Finally, we mention that there are some important superionics (type III)[60], such as β-alumina, for which the immobilized ions are fixed into a covalent network of chemical bonds. This network structure serves to further stabilize the crystal structure, enhance the temperature range over which superionic transport is exhibited. [61,62]

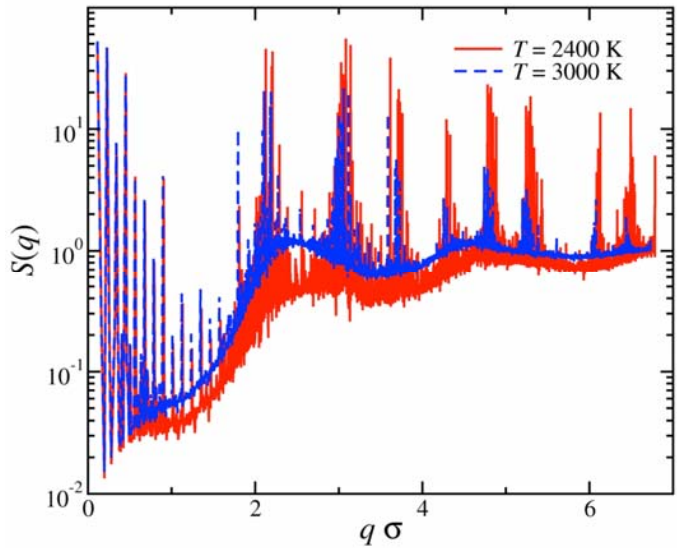

**Figure 4.** Static structure factor $S(q)$ at $T = 2400$ K and $T = 3000$ K.

Structural fluctuations near $T_\lambda$ can also be illuminated by calculating the average hydrostatic pressure exerted by the O and U ions in the crystal. The hydrostatic pressure, $P = -(\sigma_{xx} + \sigma_{yy} + \sigma_{zz}) / 3$ was calculated by averaging local atomic stress tensor for O and U ions for at least 1 ns, respectively. Figure 6 shows that the O ions exhibit a dropping pressure upon heating while the isotropic interatomic stresses felt by the U ions caused by the increased amplitude thermal motion of the O ions causes the pressure of the U ions to correspondingly increase progressively. The differential change in the U and O pressures exhibits extrema near $T_\lambda$ where the interpenetrating U and O lattices exhibits some type of structural reorganization to accommodate the high internal stresses of the ions rather than undergoing the crystal melting.

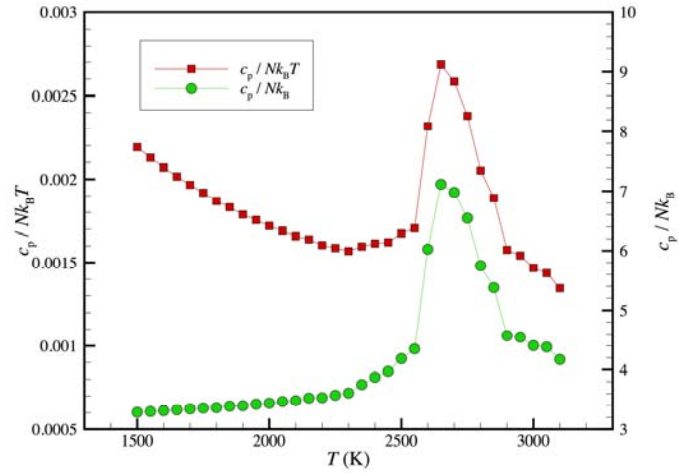

**Figure 5.** Constant pressure specific heat as a function of temperature. The position of the peak in $C_p$ defines the lambda transition, $T_\lambda$.

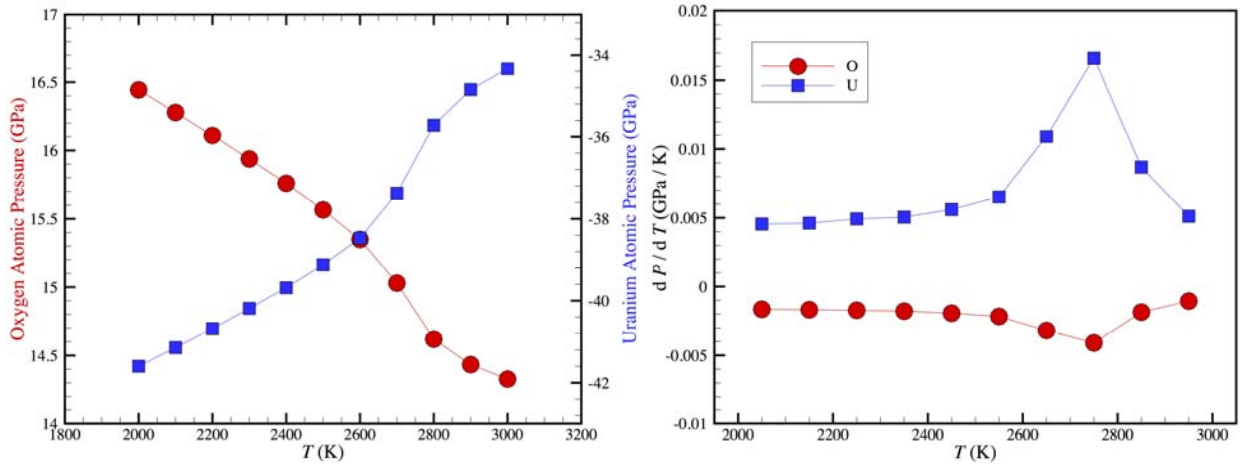

**Figure 6.** (a) The internal hydrostatic pressure of O and U ions as a function of temperature. (b) The derivative of pressure with respect to temperature for O and U ions.

Voronin and Volkov [63] have given a rather comprehensive survey of the changes of the temperature dependence of the conductivity that arise at the characteristic temperatures $T_\alpha$, $T_\lambda$ and $T_m$ of classical binary alkaline-halide class of superionics ($CF_2$, $SrF_2$, $BaF_2$, $PbF_2$ and $SrCl_2$), first discovered by Faraday (see Chadwick [4] for an accessible historical discussion of Faraday's discoveries) and Hull et al. [64] provide a thorough review of the structural properties of these crystalline materials. The superionic transition of crystalline $CaF_2$ is especially similar to $UO_2$ [65] and this material has the experimental advantage of exhibiting superionic transport at much lower temperatures. Fortunately, this situation is also true of Li salt materials that are of great interest in solid-state batteries and other applications.[30-32] Hull [66] has also provided a useful general review of superionic materials and their properties. Yakub et al. [67] and Lunev and Tarasov [68] summarize other thermodynamic and dynamic properties of $UO_2$ that are not covered in the present work such as thermal conductivity, the coefficient of thermal expansion. All these properties exhibit evidence of some kind of rounded transition resembling a second order phase transition near $T_\lambda$. We suggest that this transition is associated with dynamical disorder arising from thermally-induced structural fluctuations in the heated crystal that drive the transition from being first order to a transition more like a second order transition [57], albeit a rounded thermodynamic transition. [56] Elsewhere, this situation has be argued to also apply to cooled liquids, [20] but there is no general consensus on the thermodynamic nature of the glass transition or even if any kind of thermodynamic transition exists. As a practical aside, crystalline $UO_2$ exhibits conspicuous changes in its mechanical properties ('creep'), textural appearance and fission gas release properties near $T_\lambda$, property changes so significant that $T_\lambda$ was initially identified as the melting temperature. [58] This is just one of many examples of crystals exhibiting highly anharmonic interaction that lead to a solid-solid thermodynamic transition to an "entropically stabilized" form of the crystal where the transition has been confused with ordinary crystal melting. [69-72] There are many crystalline materials exhibiting the highly anharmonic character of $UO_2$, but which are not superionic.

In the present paper, we strictly avoid the low temperature regime below $T_\alpha$ in our simulations below. The properties of superionic materials in this "extrinsic" temperature range [4, 73, 74] are often found to depend on the history of sample preparation, the evolution of sample properties over time, and other non-equilibrium effects or "aging". These non-equilibrium effects are familiar from glass-forming liquids well below their glass transition temperature. It seems interesting that the same (2/3) $T_m$ criterion used to estimate $T_{Tam}$ is often taken as rule of estimate of $T_g$ of glass-forming liquids. [75, 76]

### B. Basic Dynamical Properties of $UO_2$

Now that we have established some basic structural aspects of the average static structure of crystalline $UO_2$ over a range of $T$, we then consider how this structure evolves from the experimentally measurable dynamic structure factor. There are two types of dynamic structure factors to be considered, i.e., the collective intermediate scattering function $F_c(\boldsymbol{q},t)$, which is naturally extending the definition of static structure factor to describe the time dependence of density correlations of the fluid as a whole, and the self-intermediate scattering function $F_s(\boldsymbol{q},t)$, which focuses on the dynamics of individual particles within the background fluid. These two quantities can correspondingly be measured by coherent and incoherent neutron scattering measurements[77, 78], making these properties particularly attractive for computational and experimental study. Both $F_s(\boldsymbol{q},t)$ and $F_c(\boldsymbol{q},t)$ are normalized in magnitude by $S(\boldsymbol{q})$ so that these correlation functions vary from 1 at short times to a constant and smaller value at long times where the wave vector $\boldsymbol{q}$ is often chosen to correspond to the interparticle distance since particle relaxation involves particle diffusion on a scale of this magnitude and in the present work $q = 2.1$ Å$^{-1}$. The intermediate scattering function is the Fourier transform of an important liquid state property, the van Hove self-correlation function $G_s(r,t)$ that describes the probability a particle at given initial position (taken to be the origin) diffuses a distance $r$ away after time $t$. $G_s(r,t)$ is a simple Gaussian function for molecules undergoing Brownian motion or atoms in a perfect crystal lattice, but $G_s(r,t)$ generally deviates from a Gaussian function for dynamically heterogeneous fluids and crystals. An interesting property of the Gaussian form of $G_s(r,t)$ is that its Fourier transform of $F_s(\boldsymbol{q},t)$ is also a Gaussian function in $\boldsymbol{q}$.

Since the Gaussian form of molecular displacement function provides a basic metric of material homogeneity from a dynamical standpoint, we first provide a measure of the extent to which our material system deviates from this ideal behavior. Such non-ideal systems are termed as "dynamically heterogeneous" and the standard metric for this property is to consider the non-Gaussian parameter $\alpha_2(\Delta t)$, defined in terms of the moments of the mean square displacement $r$ of the atoms in time interval, $\Delta t$ measured from the first instant observation,

$$\alpha_2(\Delta t) = \frac{3 < r^4(\Delta t) >}{5 < r^2(\Delta t) >^2} - 1 \qquad (1)$$

a quantity that equals 0 if $G_s(r,t)$ is Gaussian. We see from Figure 7 that the diffusion of O ions in $UO_2$ is certainly not Gaussian, and this quantity similarly to glass-forming exhibits a maximum at a time $t^*$ which grows upon cooling where both $\alpha_2(t^*)$ and $t^*$ also grows upon cooling. This striking similarity of the dynamics of $UO_2$ under superionic conditions was first recognized by Madden and coworkers.[79, 80]. We may get a better insight into the nature of the atomic motion of the O ions associated with the growing dynamic heterogeneity, i.e., growing $\alpha_2(t^*)$ by considering $G_s(r,t^*)$ as a function of $T$. Figure 8 (a) exhibits a multiple peaked structure at $r$ values that are corresponding to multiples of the O ion interatomic spacing. The ions are preferentially hopping between sites having multiples of the interparticle distance, a behavior observed in many glass-forming liquids as well where there is no lattice structure, but rather preferred scales of molecular packing exhibited by $g(r)$ that define this atomic hopping scale.

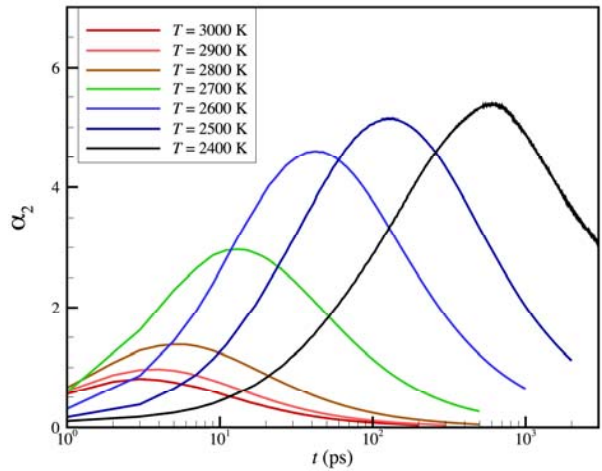

**Figure 7.** The non-Gaussian parameter $\alpha_2$ of O ions at different temperatures. The peak value $\alpha_2(t^*)$ increases monotonically upon cooling.

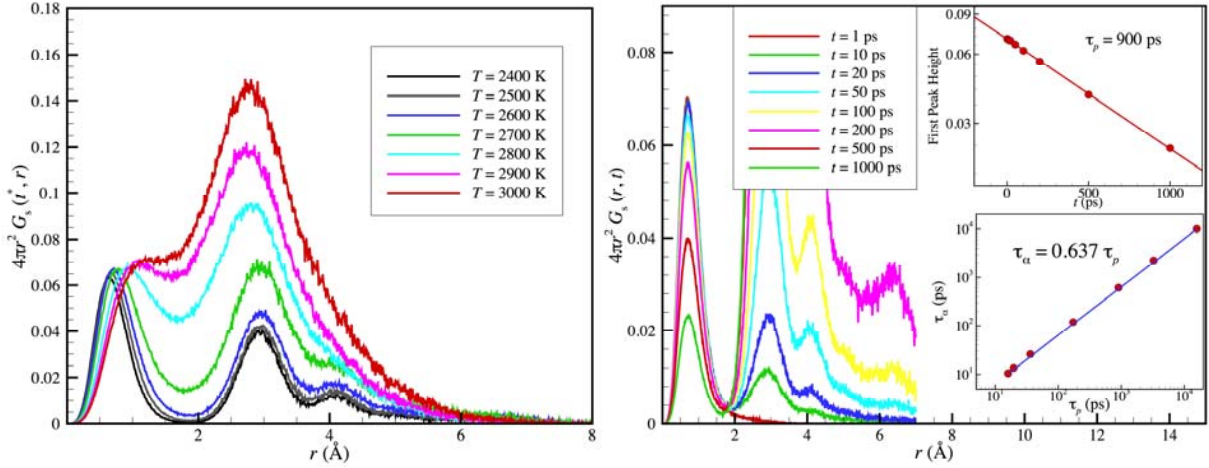

**Figure 8.** (a) Self- van Hove correlation function $G_s(r,t^*)$ of O ions at different temperatures at the time where non-Gaussian parameter $\alpha_2$ shows maximum. (b) Self- van Hove correlation function $G_s(r,t)$ of O ions for different time interval at $T = 2600$ K. The upper inset shows the first peak intensity decays exponentially, yielding the average decay time of the peak $\tau_p = 900$ ps at $T = 2600$ K. The lower inset shows the average decay time of the peak $\tau_p$ linearly relates to the relaxation time $\tau_\alpha$.

We obtain a different view of this hopping phenomenon in Figure 9 where we show the Fourier transform $F_s(q,t)$ of $G_s(r,t)$ as a function of $t$ in which again we take $q = 2.1$ Å$^{-1}$. At short times there is a initial decay associated with the fast inertial motion of the particles and then there is a longer time relaxation associated with the motion of atoms trapped in their cages, whose size is set by a transient plateau in $F_s(q,t)$. At long times, $F_s(q,t)$ decays nearly exponentially in time, a behavior quite unlike the typical phenomenology of cooled liquids where stretched exponential relaxation of $F_s(q,t)$ with a $T$ dependent relaxation stretching exponent $\beta < 1$ is normally observed. This observation is highly reminiscent of previous simulations of superheated crystalline Ni [24] where $F_s(\boldsymbol{q}, t)$ was found to a good approximation to exponential. Onuki and coworkers [81] also found this type of relaxation in a model two-dimensional superheated crystal and Gray-Weale and Madden found this type of near exponential relaxation, i.e., $\beta = 1.00 \pm 0.03$, in previous simulations of a model superionic material. [79]

All these observations are consistent with the hypothesis of Zhang et al. [24] that superionic fluids are akin to crystals in a superheated state where anharmonic interactions have stabilized the crystal so that it can exist at high temperatures. The use of periodic boundary conditions easy achieves this end as a non-equilibrium state and our crystalline $UO_2$ persists in a long-lived metastable state for temperatures up to about around 3600 K for our potential model. [37] This degree of superheating is typical for crystals with periodic boundary conditions. [82] For example, the ratio of the "mechanical stability temperature" to the equilibrium melting temperature $T_m$ has been found to be near $1.22 \pm 0.09$ for a wide range of metals (Ag, Al, Au, Cu, Mg, Mo, Ni, Ti, V, Zn, Zr) [82], compared to the ratio near 1.18 obtained from our potential model. Zhang et al. [24] did not explain previously how real superionic materials could exhibit properties similar to those of a superheated crystal.

What is the mechanism of stabilization of crystals at the reasonable atmospheric pressures? Although superheating is very difficult to achieve in metals and atomic and molecular crystalline materials with van der Waals interactions, crystal superheating is actually quite common and large in mineralogical systems (e.g., superheating quartz by over 300 K), exhibiting a combination of charge and van der Waals interactions [83] and is also observed experimentally in plasmas of charged particles. [84] Recent simulations of this remarkable phenomenon have been obtained from molecular dynamics simulations of crystalline NaCl [85] where the high cohesive interaction and charge correlations near the interface of crystalline NaCl associated with the electroneutrality condition render this material to have exceptional anharmonic stabilization of the crystalline state. Increasing the pressure to hundreds of GPa will also serve to stabilize crystalline materials, as suggested for the crystalline iron in the earth's core [5-7] and the $MgSiO_3$ perovskites comprising the dominant material of the earth's lower mantle [12], but pressure stabilization of crystals is not normally available for practical applications.

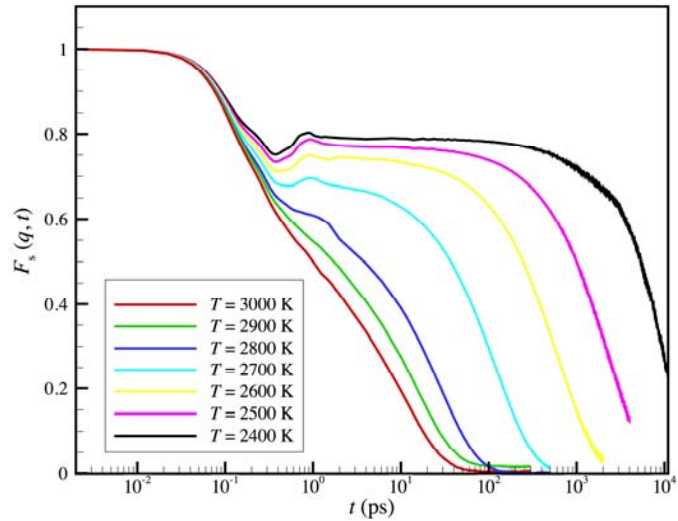

**Figure 9.** Self-intermediate scattering function $F_s(\boldsymbol{q},t)$ of O ions at different $T$ for scale corresponding to an average inter-ion distance, $q = 2.1$ Å$^{-1}$.

The rate of relaxation $\tau$ obtained from the decay of $F_s(\boldsymbol{q},t)$ can be directly related to the rate of molecular diffusion. To determine the self-diffusion coefficient $D_s(O)$ of the O ions, for example, we calculate the mean square displacement $<r^2(t)>$ of these ions as a function of time as illustrated in Figure 10(a), where we obtain $D_s(O)$ from the slope of these curves [$D_s(O)$ is determined by the long-time limit of the ratio $<r^2(t)> / 6\ t$.] We show the corresponding Arrhenius plot of our estimates of $D_s(O)$ for a range of $T$ encompassing the superionic dynamics regime in Figure 10(b). The reduction of the effective activation energy at elevated temperature evident on Fig. 10 is a characteristic feature of most superionic materials, both amorphous and crystalline, where this feature is normally reported in connection with conductivity data. [86-89] Below we show that this change in activation energy can be interpreted in terms of a change of the degree of collective motion of the O ions in the lattice.

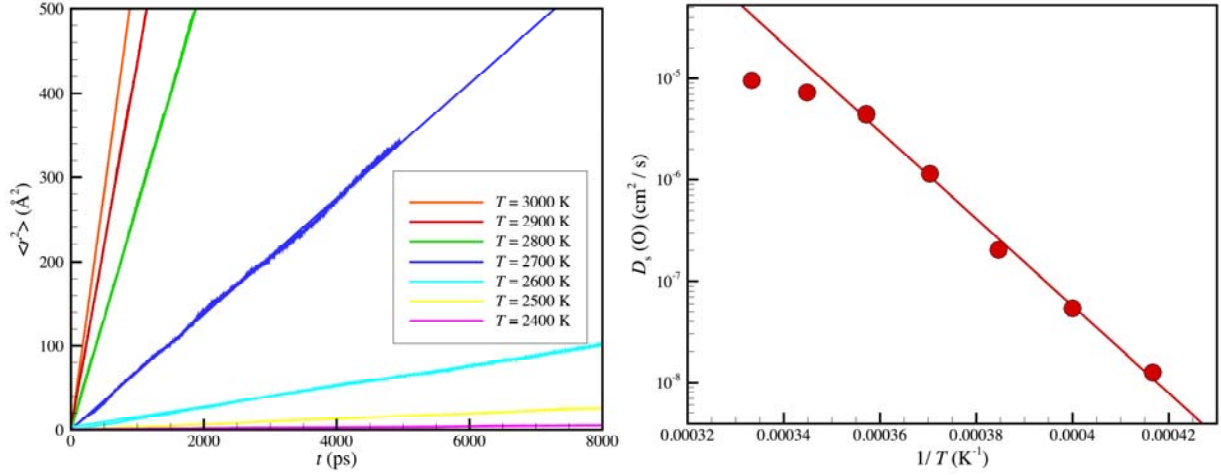

**Figure 10.** (a) Mean squared displacement of O ions as a function of simulation time at different temperatures. (b) Arrhenius temperature dependence of O ions self-diffusion coefficient.

We then plot our $D_s(O)$ estimates against our independent estimates of the relaxation time $\tau_\alpha$ obtained from fitting $F_s(q,t)$ ($\alpha$ relaxation) in Figure 11 at long times to an exponential function, $F_s(q,t) \sim \exp(-t / \tau_\alpha)$. The fitted data indicates that $D_s(O)$ scales as an inverse power of $\tau_{,\alpha}$

$$D_s(O) / T \sim (1/\tau_\alpha)^{1.04} \quad , \tag{2}$$

a relation reminiscent of the Stokes-Einstein relation between the self-diffusion coefficient and the structural relaxation time of a fluid. We observed this relation before in the simulation of superheated crystalline Ni. [24] In cooled liquids, the magnitude of the scaling exponent in this type of relationship is normally significantly less than unity, a phenomenon termed

"decoupling", but in our superionic crystalline material this exponent seems to be *larger* than unity if the statistical significance of our exponent estimate is accepted at face value, a behavior never seen in glass-forming liquids. Decoupling would appear to be operating in the opposite direction in superionic crystalline materials. There have been experimental reports of this unusual type of $D_s$ - $\tau_\alpha$ scaling in real superionic materials [90, 91]

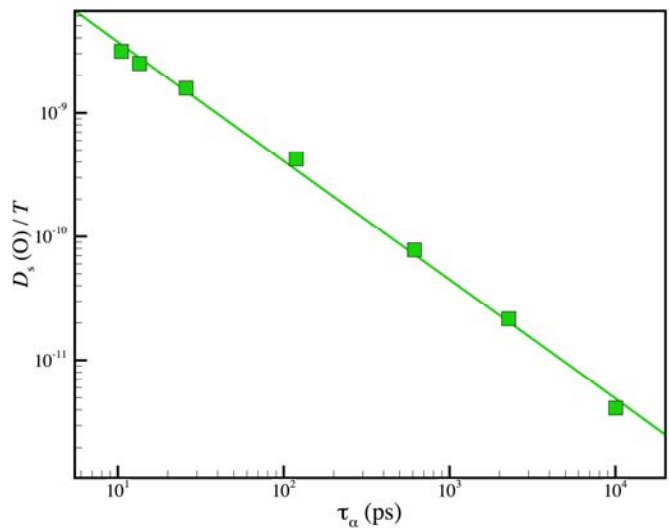

**Figure 11.** The correlation between self-diffusion of O ions obtained from mean squared displacement and relaxation time obtained from self-intermediate scattering function.

The relation between $D_s$ and $\tau_\alpha$ in Fig. 11 can be understood from a single particle perspective as arising from a hopping process of the O ions between adjacent lattice positions. This interpretation basic properties governing relaxation and diffusion in the superionic material is made apparent by examining the decay of height of the first peak of the van Hove function in Fig. 8(b) in time at $T$ = 2600 K, this peak reflecting the localized rattling of the O ions at times before they escape their cage. The upper inset to Fig. 8(b) shows that the height of this peak decays exponentially in time and the lower inset shows that the average decay time of the peak $\tau_p$ is proportional to $\tau_\alpha$ to a an excellent approximation, $\tau_p = 1.57 \ \tau_\alpha$, so that $D_s / T$ scales inversely to the average "hopping time" $\tau_p$, i.e., $D_s / T \sim \tau_p$.

To confirm that we are really simulating a crystalline material exhibiting a mobility akin to a simple fluid that decays to zero after a structural relaxation time $\tau_\alpha$, we calculate the collective intermediate scattering function, which should not decay to zero at long times in a crystalline state. Figure 12 compares $F_s(\boldsymbol{q},t)$ and $F_c(\boldsymbol{q},t)$ for O ions over a range of $T$ and $\boldsymbol{q}$ values. We observe that $F_c(q,t)$ plateaus to a constant value at long times which becomes lower in magnitude as the temperature becomes higher while $F_s(\boldsymbol{q},t)$ decays to zero at long times. We have seen this effect before in our previous studies of homogeneous melting in crystalline Ni in the pre-melting regime. Again, we have evidence that our superionic material has some resemblance to a superheated crystalline material.

We further note that the O ion $F_c(\boldsymbol{q},t)$ develops an unusual oscillatory tendency for $q \ \sigma = 2.1$, as illustrated Fig. 12(c), and then appears to plateau at long times, reflecting the high stability of the lattice of U atoms below $T_\lambda$. This oscillatory feature is highly reminiscent of the Boson peak feature of glass-forming liquids [92-94], the interfacial dynamics of nanoparticles [95] and superheated crystals [24], but the amplitude of the oscillations seems to be much larger in our superionic material. A Boson peak feature has been observed to be a universal and conspicuous feature of both crystalline and non-crystalline superionic materials. [96-98] Recent incoherent inelastic neutron scattering and accompanying simulations of high density amorphous ice have identified this type of localized low frequency oscillatory motions as taking the form of chains of molecules exhibiting collective oscillatory motion, [99] and we expect a similar phenomenon underlies these collective oscillations in our superionic material. (We illustrate the Boson peak for $UO_2$, calculated in the conventional fashion, in Supplementary Information). As the temperature approaches $T_m$, the oscillations become progressively washed out as the U atom lattice also begins to lose its stability. Local collective oscillations figure prominently in the "bond strength–coordination number fluctuation model" of Okada et al. [87] and we suggest that the highly correlated oscillatory motion evidenced by the collective intermediate scattering function at a scale of the interatomic distance may play a role in the larger scale collective hopping process we observe associated with diffusion.

### C. Collective Atomic Motion in Condensed Materials

A large number of recent computational studies of the slowing down of the dynamics of cooled liquids under equilibrium conditions has revealed that strong continuous change in the dynamics is accompanied by the growth of string-like collective motions whose extent grows upon cooling where the average length of these dynamical polymeric structures directly determines the change in the activation free energy governing mass diffusion and structural relaxation. [19, 28, 29, 100] It is then natural to inquire if the diffusion and relaxation from $F_s(q,t)$ can be described based on this same theoretical framework.

Before beginning our analysis of collective motion in superionic crystalline materials, we must point out that numerous previous simulation studies have indicated that this collective motion involves the exchange motion of mobile atoms of one type moving in a background of far less mobile atoms, which in the present instance corresponds to O ions moving with a background interpenetrating lattice of U ions. An examination of the mean square displacement of the O and U ions on a picosecond timescale, which defines the Debye-Waller factor of the respective ions clearly confirms this as a reasonable physical picture of $UO_2$ over a wide $T$ range (See Supplementary Information). Some early work suggested that the motion of the ions resembles the motion of a caterpillar where the movements occur sequentially [101], while a later computational study by Wolf [102] indicated that the cooperative motion in a simulated LiN superionic material occurred as a collective 'sliding' of a whole group of atoms at the same time, forming a more or less random particle exchange path. This phenomenon in superionic crystals was interpreted in terms of a solitary wave conduction model of transport in these materials proposed earlier by Bishop.[103] Zheng-Johansson and Mcgreevy [104] also correlated motions of chain moving in a collective fashion in a simulated superionic CuI material where interesting illustrations of representative paths are given. Simulations on glass-forming liquids has indicated that this exchange motion varies between these extreme limits depending on conditions so there is probably some truth to both these models of collective motion in superionics. Recent work has further confirmed this collective exchange motion in Li salt materials is in connection to solid-state batteries. [31, 105] Eapen and coworkers [25-27] have also examined this kind of collective motion recently for a range of type II superionic materials ($CaF_2$, $UO_2$, $LiO_2$) where it was suggested that ionic conductivity in these materials is governed by the lifetime of the string like ion clusters defined by a time at which the number of particles in string clusters become maximized. While this work unequivocally establishes the existence of string-like excitations in these materials and provided much needed quantification of the average properties, the agreement of the model predictions with simulation data is only qualitative. The analysis also does not accord with analyses performed previously on glass-forming liquids where the significance of the strings is related to their relation to the temperature dependence of the activation energy governing mass diffusion and structural relaxation. We revisit the strings in $UO_2$, adopting the standard approach applied previously to glass-forming liquids.

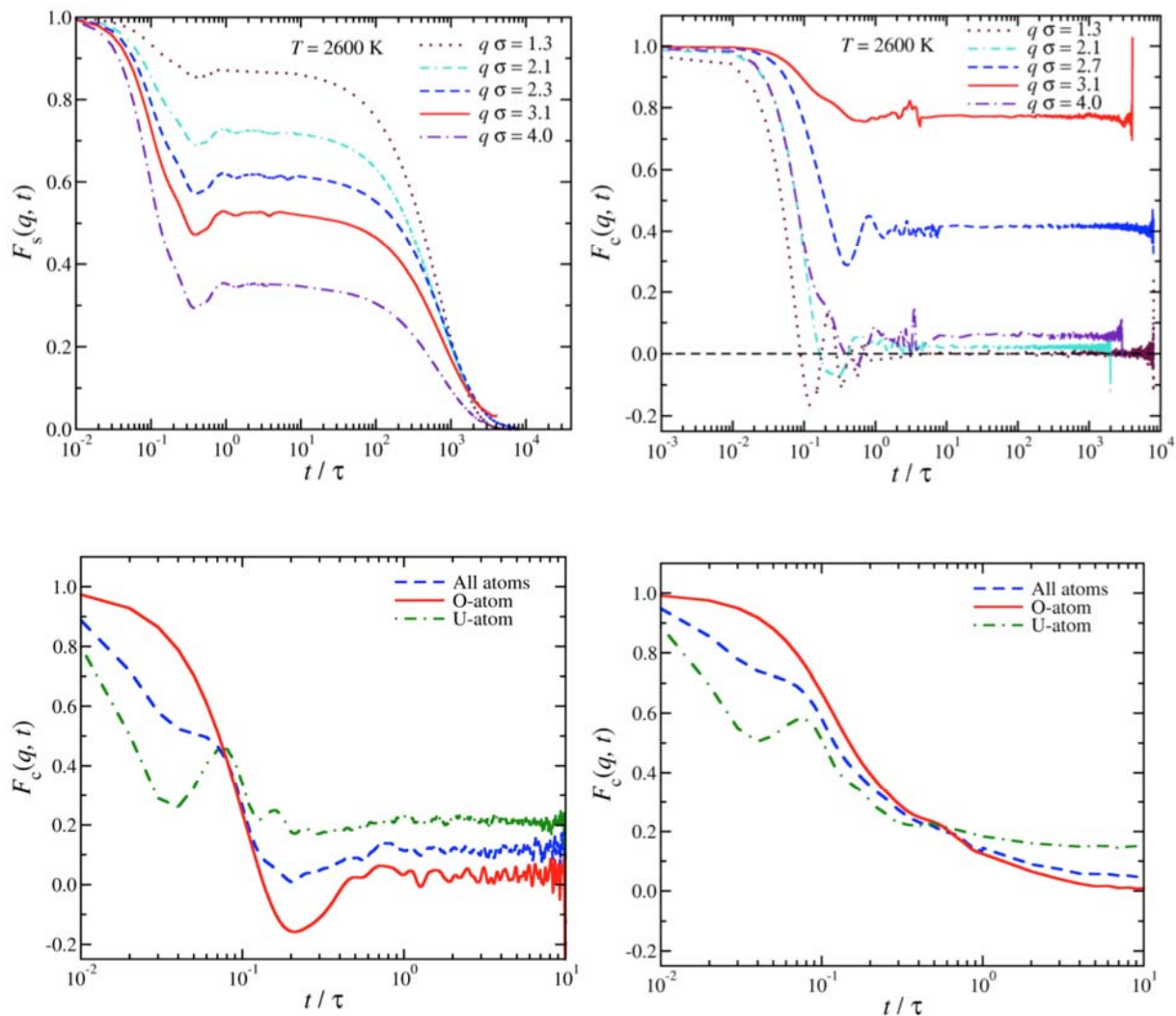

**Figure 12.** (a) Self- and (b) collective- intermediate scattering function of O ions at $T$ = 2600 K for a range of $q$ values. Collective intermediate scattering function of U atoms, O atoms, and all atoms at (c) $T$ = 2500 K and (d) $T$ = 2900 K.

### D. Quantifying the String-Like Collective Motion in $UO_2$

Cooperative particle dynamics is one of the most characteristic features of the dynamics of GF fluids. Both atomistic simulations and experiments on glass-forming colloidal and granular fluids exhibit string-like collective motion and MD simulations have recently shown that this type of motion also occurs in the dynamics of grain boundaries[106] and the interfacial and melting dynamics of NPs.[21, 107] We next apply methods originally developed to identify this type of correlated motion in GF liquids to our simulations of the dynamics in $UO_2$. First, we review the definition of collective atomic displacement motion.

As a first step in identifying collective particle motion, we must identify the "mobile" atoms in our system. In GF liquids, the "mobile" atoms (atoms with enhanced mobility relative to Brownian motion) are defined by comparing the self-part of the van Hove correlation function $G_s(r, t)$ for the strongly interacting particle fluid to an ideal uncorrelated liquid exhibiting Brownian motion where $G_s(r)$ reduces to a simple Gaussian function. Since "mobile" atoms are essentially those particles moving a distance $r(t)$ larger than the typical amplitude of an atomic vibration after a decorrelation time $\Delta t$, but smaller than the second nearest-neighbor atomic distance (as suggested by the multiple peaks of the self-part van Hove correlation function shown in Figure 8), we mathematically identify these mobile particles, as in previous studies of GB dynamics[106] and the interfacial dynamics of NP by a threshold atomic displacement condition[107], $a < |\ \mathbf{r}_i(\Delta t) - \mathbf{r}_i(0)\ | < b$, where constants $a$ and $b$ that can be determined from the van Hove correlation function. The further identification of collective atom motion requires a consideration of the relative displacement of the particles. Collective atomic motion implies that the spatial relation between the atoms is preserved to some degree as the atoms move. In particular, the reference mobile atoms $i$ and $j$ are considered to be within a collective atom displacement string if they remain in each other's neighborhood, and we specify this proximity relationship by, min $[|\mathbf{r}_i(\Delta t) - \mathbf{r}_j(0)|, |\mathbf{r}_i(0) - \mathbf{r}_j(\Delta t)|] < 1.1$ Å. Atomistic simulations of GF liquids indicate that the distribution of string lengths $P(n)$ is approximately an exponential function of the number of atoms in the string $n$,

$$P(n) \sim \exp(-n / <n>), \tag{3}$$

where $P(n)$ is the probability of finding a string of length $n$ at the characteristic time, $t^*$.

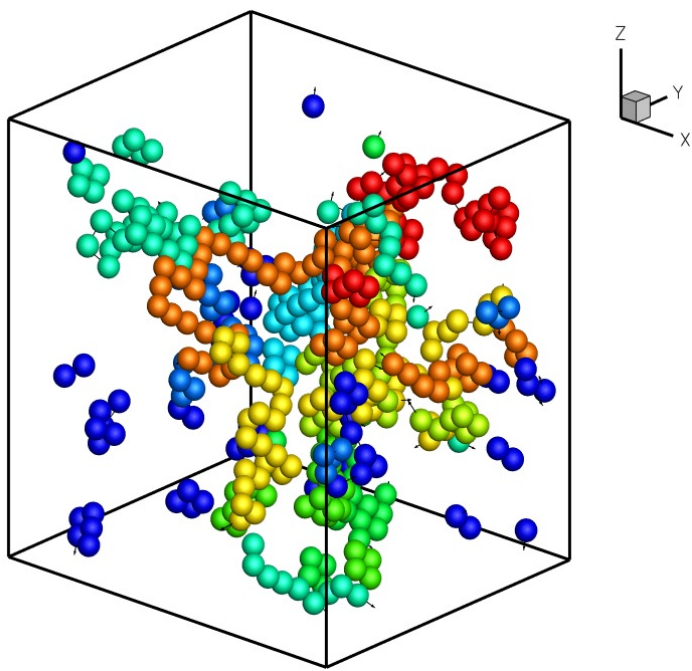

**Figure 13.** String-like collective atomic motion of O ions in UO$_2$ at $T$ = 2400 K. The colors are introduced to discriminate between different string events.

In Figure 13, we show a representative configuration of string-like exchange events among the O ions for UO$_2$ at $T$ = 2400 K determined by the methods just described. The structural configuration of these strings remarkably resembles strings identified in both metallic glass and polymeric glass-forming liquids [28, 29]. As observed in our previous study of melting of crystalline Ni, the string-like motion involves a collective hopping motion in the form of ring like particle exchange structures at low temperatures and a topological transition to open strings at high temperatures. Figure 14 shows our quantification of this gradual transformation of string topology in our superionic crystal where we see that the ring structures become exhausted around $T_\lambda$. Most of the strings take the form of ring strings as $T$ approaches the onset temperature $T_\alpha$. The strings are clearly polymer-like in form and polydisperse in length, as found before repeatedly in glass-forming liquids, [19, 28, 29, 100] and we determine the length distribution of the strings and their fractal geometry. The resulting size distribution of the strings found in our equilibrium $UO_2$ samples is indicated in Figure 15 where we find that the distribution above $T_\lambda$ is exponential to a good approximation, again as generally found in all glass-forming liquids studied to date. At temperatures below $T_\lambda$, however, we see something new. The size distribution appears to follow two exponential distributions for string lengths above and below a string length of about 10. The average value of the strings in the superionic material is notably much longer than observed typically in glass forming where the average string length only grows to a length of about 4 or 5. [108, 109] ('cooperative rearranging region size') only grows to a length between 3 to 7. [108-111] The inset to Figure 15 shows that $L$ in our cooled $UO_2$ grows to a value as large as 10 so these materials exhibit much more cooperative atomic motion than found in glass-forming liquids. This is frankly a surprising finding to us. The lattice somehow enhances the propagation of collective atomic motion in the form of randomly shaped strings rather than suppressing this motion. Again, following previous treatments of glass-forming liquids [24, 29], we examine how the radius of gyration of the strings $R_g$ scales with the number of segments $n$ in the strings. The data is somewhat noisy, but a best fit to the data over a range of temperatures indicates that the average value of the mass scaling exponent $\nu$, defined by $R_g \sim n^\nu$ caries from about 0.6 to 0.5 as the temperature is lowered, exactly as found before for glass-forming polymer liquids. [29] We conclude that the form of string excitations in superionic $UO_2$ follows the standard phenomenology of an equilibrium polymerization process for which the strings at high temperatures exhibit the geometry of as suspension of swollen polymers (self-avoiding walks), but as these structures grow and interpenetrate at lower temperatures becoming less swollen due to their greater mutual interaction, resembling random walks. Exactly the same crossover arises in solutions of equilibrium polymers so this general trend is quite understandable when these excitations are taken to be equilibrium polymer structures. [20] Of course, the whole point of quantifying the string 'dynamic heterogeneity' is to establish a relation to $D_s(O)$ and $\tau$ to the structure of these fluid excitations and we next test whether the same relations as found in glass-forming liquids are applicable to our superionic material.

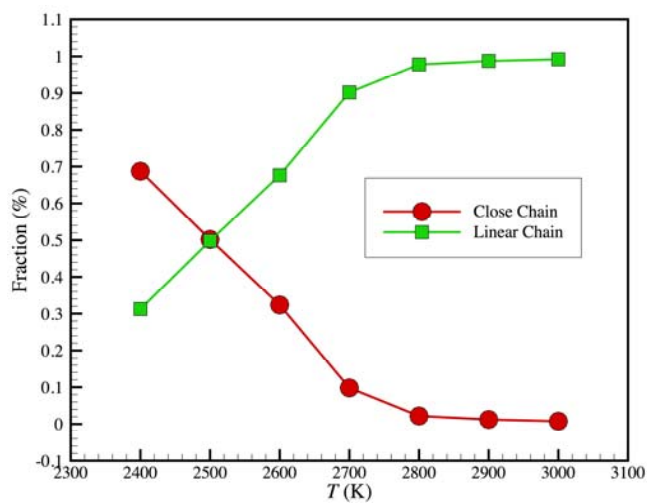

**Figure 14.** Fraction of open (linear) strings and close strings as a function of $T$.

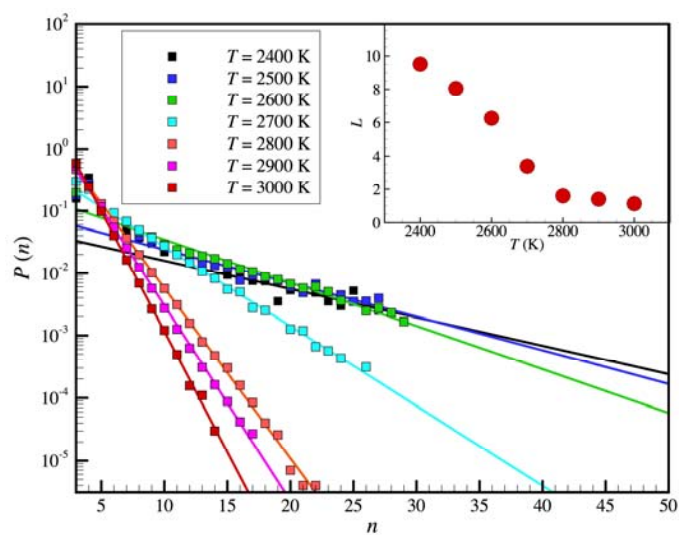

**Figure 15.** The string size distribution at different temperatures. Inset shows the $T$ dependence of the average string length, $L = <n>$.

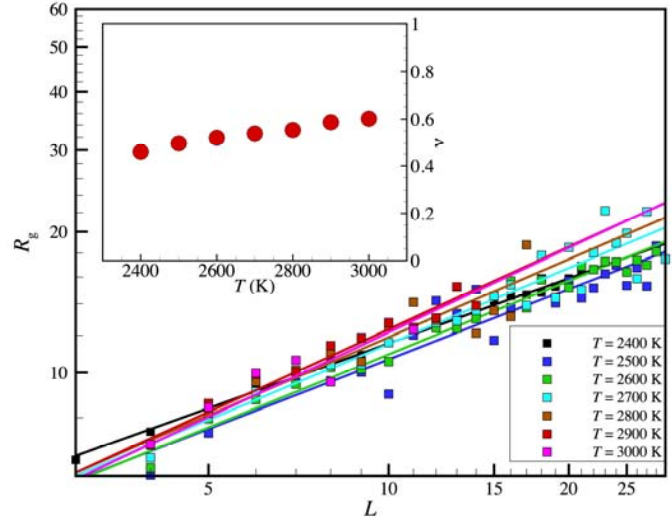

**Figure 16.** Scaling of string radius of gyration $R_g$ with its length $L$ as $R_g \sim n^\nu$, where the inset shows $\nu$ as a function of $T$.

### E. Test of the String Model of Relaxation to $D_s(O)$ and $\tau$ of a Model Superionic Material

We are now in a position to determine if the length of the strings $L$ is directly related to the temperature dependence of the activation free energy for molecular transport, as found in glass-forming liquids. Confirmation of such a relationship would potentially greatly expand the realm of application of this model of relaxation and dynamics to include crystalline materials. The possibility that such a relation might hold is suggested by the previous successful application of the Adam-Gibbs model of the dynamics of glass-forming liquids to the dynamics of superionic materials in simulations by Gray-Weale and Madden.[79] According to the string model, the relaxation time $\tau$ scales as [108, 112, 113],

$$\tau_\alpha(T) = \tau_0 \exp\left[\frac{(L(T)/L_A)\Delta\mu}{k_B T}\right] \tag{4}$$

where $L_A$ is a residual collective motion at the onset temperature, in which $L$ starts to vary appreciably with temperature in glass-forming liquids (We take in the current study of superionic $UO_2$). $D_s(O)$ scales inversely to $\tau$ to a good approximation in our simulations so we confine our attention to $D_s(O)$ in our subsequent analysis. Specifically, we expect from the string model that $D_s(O)$ should be described by the relation,

$$D_s(O) = D_{s0}(O) \exp\left[-\frac{L(T)G_0}{k_B T}\right] \qquad (5)$$

where $D_{so}(O)$ and $G_o$ are taken as adjustable parameters that can be fixed in the present situation by matching this equation to $D_s(O)$ at the melting point and the temperature dependence of $D_s(O)$ for temperatures near $T_m$ where $L$ is slowly varying. We compare Eq. (5) to our $D_s(O)$ in Figure 17 and find that the agreement is quite reasonable. Following previous experimental studies,[87] we may also fit our $D_s$ data reasonably well to the Vogel-Fulcher-Tammann (VFT) expression [1] that is often used as an empirical fit function for diffusion coefficient and relaxation time data in glass-forming liquids. However, this functional form does not capture transition to an Arrhenius relaxation at low $T$ in the superionic mobility [$D_s(O)$] data, as observed also in experiments on crystalline superionic materials. [114, 115] A corresponding transition from VFT to Arrhenius relaxation at low temperatures has also been reported in glass-forming liquids. [116-119]

Previous simulation work by Gray-Weale and Madden [79, 80] likewise recognized the strong analogy between certain superionic crystalline materials and glass-formation in cooled liquids and these authors were able to successfully describe their $D_s(O)$ and $\tau_\alpha$ simulation data for model alkaline halide superionic crystalline materials in terms of the Adam-Gibbs model of relation in glass-forming liquids where $\tau_\alpha$ is proposed to scale as $\tau_\alpha \sim \exp(A/S_c T)$ in which the configurational entropy $S_c$ of this model was estimated from the excess entropy of the crystal minus its value at $T_\lambda$ and $A$ is a constant related to the activation energy parameter $\Delta\mu$ in Eq. (4).

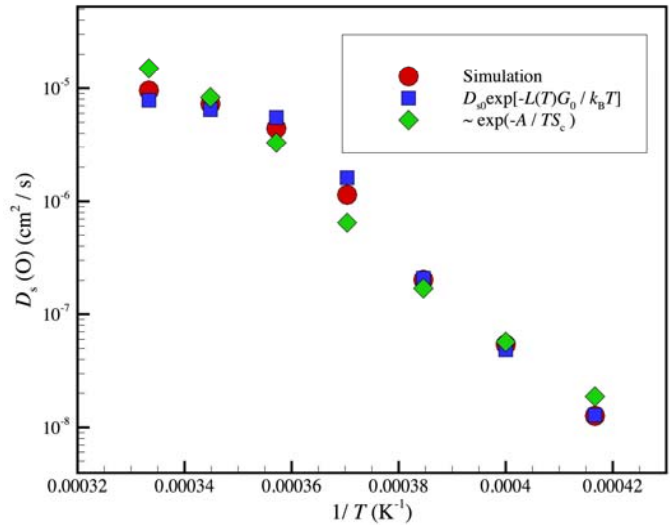

**Figure 17.** Fitting of $D_s(O)$ data to string model of relaxation and the AG expression for $\tau$, where $S_c$ is approximated by the excess entropy density estimate.

As discussed elsewhere, the string model result is consistent with the AG model, where $L$ is to be exactly identified with the abstract cooperatively rearranging regions of the AG model. In the absence of a direct calculation of $S_c$ for our crystalline material, we follow a procedure similar to that employed in previous studies of glass-forming liquids [120] a procedure that differs somewhat from the previous test of the Adam-Gibbs relation for superionic $UO_2$ by Gray-Weale and Madden. In particular, we first estimate an *excess entropy $S_{exc}$* as difference between the material entropy and the vibrational entropy of the low temperature entropy of the material arising from vibrational entropy of the lattice as an estimate of the configurational entropy of the lattice, a procedure similar to Richert and Angell, [120] where the vibrational entropy of the liquid is subtracted from the total fluid entropy based on the assumption that the vibrational entropy of the crystal approximates that of the fluid. As in the calculations of Gray-Weale and Madden, we estimated the total entropy $S$ of the superionic material based on a determination of the specific heat by simulation (See Figure 5) and we then estimated the vibrational contribution by assuming $S$ is entirely vibrational below the onset temperature $T_\alpha$ where anharmonic effects are weak. The difference of $S$ minus its value below $T_\alpha$ then defines $S_{exc}$, which should reasonably approximate $S_c$ of the heated crystalline material. As a further refinement of this type of AG modeling to describe glass-forming materials at constant pressure [121], we normalize $S_{exc}$ by the material volume $V$ so that our barrier height is governed by excess entropy density. The AG model at constant pressure gives unphysical results [121] at elevated temperatures without this refinement since it never saturates at elevated temperatures and in Figure 18 we compare the excess entropy per unit mass and per unit volume estimated in the way just described from our $UO_2$ specific heat and thermal expansion data. The excess entropy density estimates clearly have the sigmoidal variation with temperature that qualitatively conforms to the $L$ variation determined before for this material, but the excess density without the volume normalization does not have this property, as found in experimental estimates of glass-forming liquids. We show the result of fitting our $D_s(O)$ data to the AG expression for diffusivity where $S_c$ is approximated by the excess entropy density estimate, illustrated in Figure 17. We see that this fitting leads again to good accord with our $D_s(O)$ data, suggesting further a scaling $L \sim 1/S_c$, a basic premise of the AG model and a finding of the string model of glass-formation. [108]

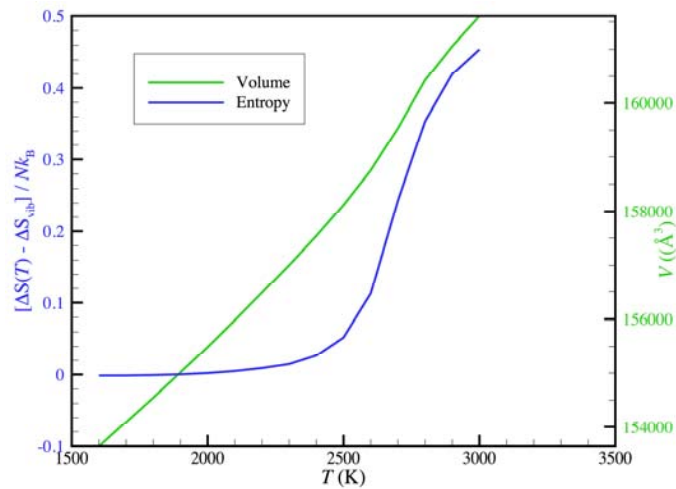

**Figure 18.** Excess entropy and volume as a function of temperature, estimated based on UO₂ specific heat and thermal expansion.

## *F. Interfacial Diffusion*

Most crystalline materials are actually polycrystalline in the sense that they contain a substantial of grains having grain boundary regions in which mobility and relaxation can be greatly different from mobilities deep within the crystal. The mobility in these interfacial regions has often been observed to many orders of magnitude larger so that the conductivity can be dominated by these interfacial regions. This phenomenon has also been observed in certain fluorite superionic materials.[122-125] In general, these changes in mobility are highly grain boundary specific and previous work on the grain boundaries of different types of Ni [106] has shown that this variability reflects the variable extent of collective motion in these different types of interfacial regions and a similar highly variable mobility has been observed in the interfacial regions defined by the different crystallographic interfaces of an isolated Ni crystal. [126] Given the importance of interfacial mobility of superionic materials in applications, we briefly consider the interfacial mobility on the (110) interface of UO₂.

As we have observed before in the case of Ni and H₂O interfacial regions, heating UO₂ of simulated crystal with a crystal-vacuum interface leads to the formation of a mobile interfacial layer in which the Debye-Waller factor is enhanced relative to the bulk material modeled with periodic boundary conditions and a greatly accelerated diffusion for temperatures above the Tammann temperature, which in the present instance is close to $T_\alpha$. We illustrate the diffusion coefficient in the mobile interfacial region, defined as before [126, 127], in Figure 19, where we see a similar overall pattern of behavior as found for the bulk self-diffusion coefficient $D_s(O)$ of the mobile species. To avoid confusion, we designate $D_s(O)$ in the mobile interfacial region as $D_s^{int}(O)$.

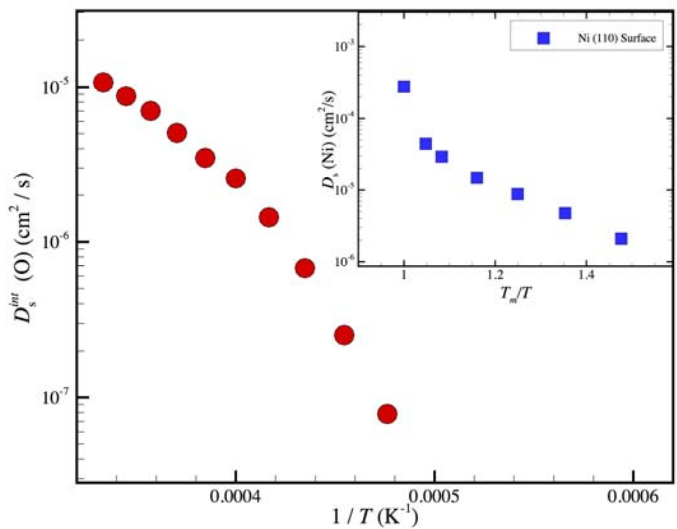

**Figure 19.** Interfacial (110) self-diffusion coefficient $D_s^{int}(O)$ as a function of temperature. The inset shows the interfacial self-diffusivity on Ni (110) surface.

We note that the ratio of $D_s^{int}(O)$ to $D_s(O)$, quantifying the interfacial acceleration of diffusion in this interfacial region of UO₂, is much smaller than observed in previous simulations of heated crystals. Tentatively, we associate this trend with the high cohesive energy of ionic versus non-ionic materials [83] (see discussion above), which can be expected to diminish the relative amplitude of atomic motion in the interfacial region and lead to a general tendency of ionic crystalline materials to superheat.[85] The relative suppression of interfacial mobility in ionic materials such as $UO_2$ due to their relatively strong self-cohesive interaction might then be intimately linked to the relative stability of this class of solid materials at elevated $T$, the relative anharmonic nature of crystals of these crystals at high $T$ and the occurrence of a dynamics that greatly resembles superheated crystals that are in non-equilibrium high $T$ metastable state.

There is another striking difference in the interfacial mobility of superionic $UO_2$ in comparison to "normal" crystalline interfacial regions typified by Ni and $H_2O$. [126, 127] At the "breakpoint" in the Arrhenius plot describing the $T$ dependence of the interfacial diffusivity, $D_s^{int}(O)$ bends down at high temperatures in the superionic material, while we have found that this quantity *bends up* in the high temperature regime for "normal" crystalline materials (Note that the breakpoint in the Arrhenius curve also in the bulk $D_s(O)$ data; See Fig. 17). For comparison, we show our previous finds for Ni in the inset of Figure 19 and note that measurements showing that this behavior is rather "universal" in the interfacial dynamics of metal films is summarized by Rhead[128], Binh and Melinon[129], and Bonzel and Latta[130]. It would appear that the interfacial dynamics is quite different in superionic materials than "normal" crystals.

Convex Arrhenius plots similar to our superionic material have been observed in the catalytic reaction dynamics of proteins where the 'kink' in the Arrhenius curve corresponds to some kind of dynamical transition from an active to a relatively inactive catalytic state. [131-133] In this context, the transition is attributed to a softening transition of the protein upon heating [134], which is signaled by an appreciable drop in the activation energy of the catalytic reaction rate in the high temperature region above the kink accompanied by a corresponding huge increase of the reaction rate prefactor (changes as large as $10^{12}$ have been observed experimentally in some proteins) that has been associated physically[132] with the proliferation of many distinct activated conformational substates in the protein. Moreover, the change of the entropy of activation corresponding to these prefactor changes is quantitatively described by an entropy-enthalpy compensation relation in which the change in the enthalpy of activation is equal to that of the activation entropy times the breakpoint temperature. This physical picture of the origin of convex Arrhenius relaxation in terms of a softening transition to a material state exhibiting significant configurational fluctuations of the atoms seems to apply equally well to our superionic material, and when we analyze the activation enthalpy and entropy of $D_s(O)$ in our superionic $UO_2$ we also see entropy-enthalpy compensation. The observation of transition to a low activation energy at high temperatures is one of the most characteristic features of superionic materials. [97]

It is natural to expect that the convex Arrhenius dynamics of protein catalysis can be also interpreted on a molecular scale in terms of a change in the collective atomic motion with $T$, as we have observed in our superionic $UO_2$ simulations. Recently, we have shown that string-like collective motion is clearly present in the internal dynamics of proteins at $T$ below the protein denaturation temperature [135], although simulations have not yet been performed over a large $T$ range to exhibit an abrupt transition in the scale collective motion as $T$ is varied. However, the string model of condensed matter relaxation leads us to generally expect that the entropy of activation should be altered in a concerted way with the activation energy so that large pre-factor changes in Arrhenius plots can generally naturally arise when changes in the scale of collective motion $L$ are abrupt. Since a large decrease of rigidity can generally be expected in a folded protein upon approaching denaturation temperature, or a superionic material upon approaching its melting temperature, these materials can both be expected to exhibit a sharp increase of configurational entropy $S_c$ and a corresponding decrease in the scale of collective motion. Efficient enzyme function of proteins and the high conductivity of superionic and other highly anharmonic crystalline materials seems to rely on some stabilizing framework that allows for the system to retain some structural integrity while at the same time a large degree of internal structural fluctuations to enable a high rate of reactivity and transport.

### G. Conductivity of Superionic $UO_2$

The conductivity $\sigma$ of superionic materials is perhaps the most significant property of these materials. Although the variation of $\sigma$ generally shows similar qualitative changes with temperature as $D_s(O)$, as expected from the Nernst-Einstein relation, we may expect deviations from the simple predicted proportionality relation estimated for these properties arising from ion correlations neglected by the Nernst-Einstein relation. In particular, we may expect the string-like correlated exchange motion to enhance $\sigma$ relative to the Nernst-Einstein relation. This is another topic deserving a separate study in view of the computational difficulty of estimating precise estimates of $\sigma$. Some discussion of this phenomenon seems necessary in any discussion of superionic materials.

In Figure 20, we compare the calculation of collective diffusion coefficient, which determines $\sigma$, and the self-diffusion coefficient from mean square displacement data obtained for our simulated $UO_2$ material at representative temperatures above and below $T_\lambda$: $T = 2600$ K and $T = 2900$ K. The details of the definition of $\sigma$ and method of computation for the collective diffusion coefficient have been described thoroughly before by Madden and coworkers[31, 136] and we do not reproduce this analysis here. We see in these examples that $\sigma$ is enhanced relative to its Nernst-Einstein value by a factor of about 2 to 3 for these examples. Much larger enhancements in $\sigma$ have been observed in simulated Li-based superionic materials of interest for battery applications.[31] The ratio of $D_s(O)$ to the collective diffusion coefficient, determined from conductivity, the "Haven ratio" [4, 137], is decreased appreciably by collective motion so that the "inverse Haven ratio" defines the extent to which collective ion interactions enhance ion conductivity. There are few precise calculations of $\sigma$ over a wide $T$ range for superionics because of the long computational times required to obtain precise numerical estimates, and these calculations are particularly difficult to perform near $T_\lambda$ where fluctuation effects are appreciable. The investigation of requires a study dedicated to this problem and here we just note that $\sigma$ varies in a parallel fashion to $D_s(O)$, but the Nernst-Einstein relation does not hold quantitatively.

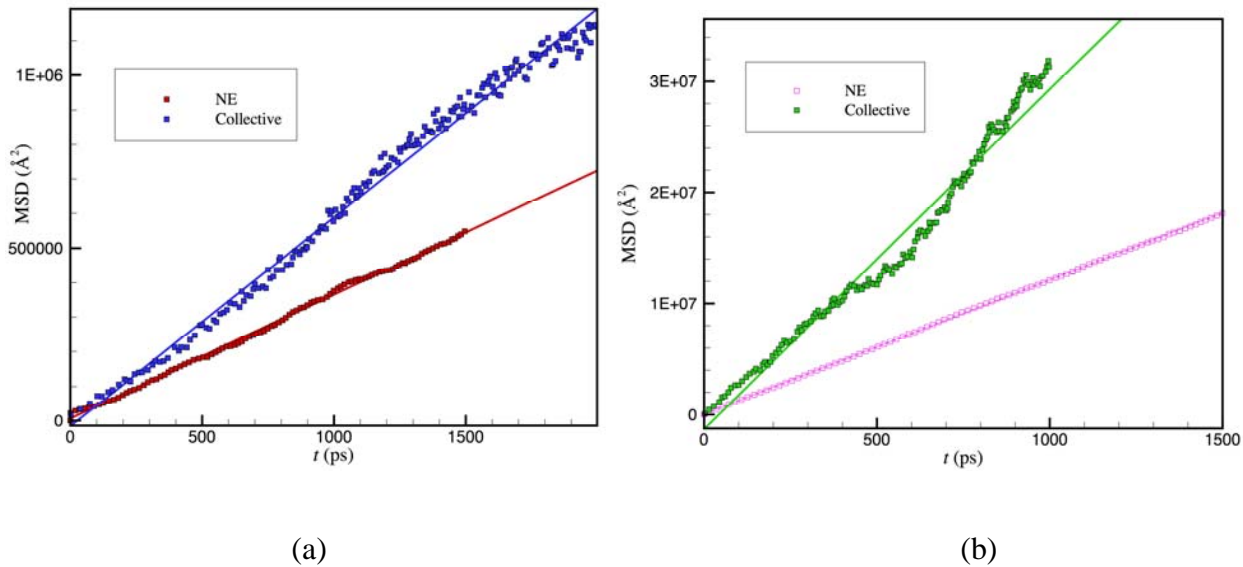

(a)                                            (b)

**Figure 20.** Comparison of collective and regular mean squared displacement of O ions at temperatures above and below $T_\lambda$: (a) $T$ = 2600 K and (b) 2900 K.

Our discussion of the conductivity raises an interesting question regarding how collective influences $\sigma$ in comparison to the ion mobility measured by $D_s(O)$. Voronin and Volkov mentioned in their comprehensive review of alkaline-halide (Type I)I superionic conductivity and thermodynamic data, [63] that the activation energy for $\sigma$ exhibits a maximum near $T_\lambda$ in all systems they investigated ($CF_2$, $SrF_2$, $BaF_2$, $PbF_2$ and $SrCl_2$), but we see in Figure 15 that the string length, and thus the activation energy, exhibits no maximum in the vicinity of $T_\lambda$ in our $D_s(O)$ data for $UO_2$. Assuming $D_s(O)$ scales linearly with $\sigma$, this suggests that there may be some essential difference in the variation of $\sigma$ with $T$ in $UO_2$ from the common Type II superionics studied by Voronin and Volkov. Unfortunately, we cannot find any $\sigma$ data for temperatures passing through the $T_\lambda$ of $UO_2$ to check if this material exhibits a maximum in its activation energy for $\sigma$.

In our previous simulations of homogeneous melting of crystalline Ni, we did observe a maximum in the string length as a function of $T$ below the mechanical melting transition. At that time, we suggested that superionic materials might be superheated materials, or a material very much like them, based on the existence of this $L$ maximum, and the corresponding maximum in the activation energy in measurement of the fluorite minerals. This suggests that we should take a closer look at the string length distribution found for our $UO_2$ system (See Fig. 15). We see from this data that the strings having a length less than the crossover value n $\approx$ 10 indeed grow non-monotonically with temperature and we plot the variation of these short strings in Figure 21, where we see that the short string length peaks near $T_\lambda$. There is actually a waxing and waning of the collective motion in this sense, as suggested by other observations Eapen and coworkers.[56, 138]

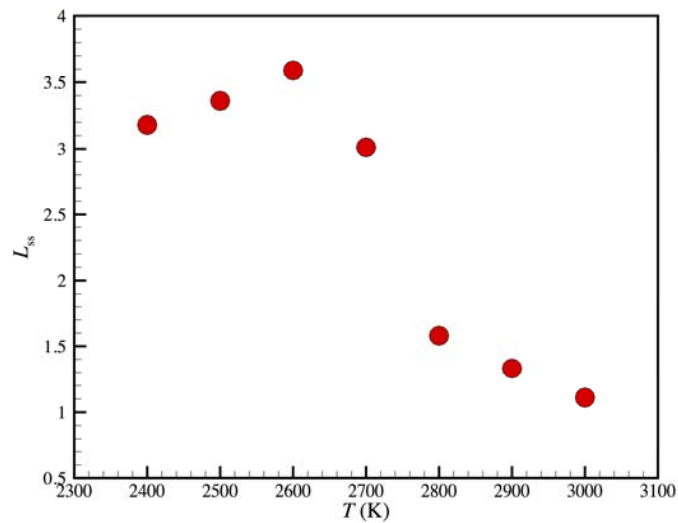

**Figure 21.** The average short string length $L_{ss}$ as a function of temperature.

We emphasize that we have never seen the string length vary in a non-monotonic way with $T$ in glass-forming liquids so how might we understand this at first counter-intuitive trend in the degree of collective motion? The string model of relaxation in condensed materials, [20, 108, 139] potentially provides some insight into this unexpected and apparently non-universal trend. In the string model, the temperature dependence and their overall length is regulated by defects in the liquid similar to interstitial defects in crystal. These proposed "defects" (apparently these defects quite observable in some liquids such as water [140] and such defects are also central to the Granato "interstitiality" model [141] of glass-forming liquids and defective crystals) play the role of an initiator in generating string excitations of a polymeric form involving cooperative molecular exchange motion - a larger concentration of these defects, the smaller the chains and weaker the $T$ dependence of the strings from the string model. Interstitials are certainly an important class of defects and have previously been shown to mediate the string-like collective motion observed in our simulations of superheated crystalline Ni. [24] An important predicted property of the "interstitiality" defects is that their concentration should be small, but constant, in liquids, but their concentration is $T$ dependent, approaching an asymptotic value in the crystal as it approaches its melting point. In a crystal, we may expect the concentrations to increase steadily, at first increasing the extent of collective motion relative to the perfect crystal at low temperatures where collective motion is largely absent. The generation of too many of these defects, on the other hand should ultimately give a rise to smaller strings, as predicted in glass-forming liquids where the concentrations of such defects is expected to be relatively constant.[108] The maximum in the string length is then reasonably rationalized by the competitive interaction of defects initiating collective motion. The model of the variation of the string length with temperature remains to be validated by direct simulation study and the validation of the existence and nature in liquids is also a problem deserving an intensive investigation.

## IV. Conclusions

We have confirmed the suggestion of Gray-Weale and Madden [79, 80] that great similarities exist between the dynamics of superionic materials and glass-forming liquids. In each case, the changes in the activation free energy for diffusion and relaxation can be quantified by changes in the average degree of cooperative motion, which is reflected in a common way of changes in the configurational entropy of the material. It is satisfying to obtain a unified description of such different classes of materials and we look forward to further testing the string model of relaxation in different classes of materials.

Given the emerging importance of superionic materials in diverse applications in the area of fuel cells, supercapitors, energy conversion, sensors [32, 34, 142-146] and the need for long lasting, fast charging and safe battery materials utilizing such materials[32, 143, 144, 147, 148], it is worth pointing out what has been learned from our analysis that is germane to the design of materials in this general class. Superionic materials arise in essentially all general classes of ionic materials [149] so what makes a material superionic under reasonable physical conditions (we exclude extreme pressures and other situations of limited interest from a commercial standpoint). In crystalline superionic materials [55, 62, 66], it appears to be necessary for the mobile ionic species to move in a background of ions of the opposite change that are immobilized in space due to their relatively high mass or chemical bonding into a network. For the mobile ions to move there must be low barriers for their activated transport between interstitial sites with the ion lattice. Open cubic lattices for the interpenetrating mobile ion lattice allow for these barriers to be lower than for close packed lattices. There is a high prevalence of superionic cubic materials to have cubic sublattices and in the fluorites of bcc lattices. Many superionic species can be found in associated with a bcc lattice structure stabilized over close packed lattices such as fcc at extremely high pressures and temperatures. Recent combinatoric studies of potential superionic materials for solid-state battery applications [30] have suggested that charge carrier lattices that have a bcc structure give rise to optimal superionic materials. There are also basic issues involving the size of the mobile ions that can be tuned to create an optimal superionic material.

Smaller ions tend to have to cross lower relative energy barriers to hop between positions in their lattice so the degree of superionicity tends to increase when the ratio of the mobile ion to that of those of the immobile ion lattice becomes smaller. [63] An effective strategy would seem to be having as much free volume or cavity space as possible and a low inter-ion interaction strength (say by tuning the dielectric constant) to minimize the activation energy for the mobile ions in the high temperature crystalline regime where the ions can exhibit streaming ion transport in the form of collectives of mobile ions, i.e., strings. Superionic materials having an impressive conductivity have been made in Li Carbene materials[150], an extreme example of using ion size asymmetry to create useful superionic materials. We note, however, that this trend is not as simple as smaller is better. Recent measurements have shown that ion diffusivity in superionic materials is sometimes optimized for ions having intermediate size, a phenomenon often observed in molecular transport in highly confined systems (the 'levitation effect') so there generally other physics beyond ion size that contribute to the activation barrier height for diffusion, which need to be considered while ion size is tuned.[62]

It should be possible to extend the same physical design rules to amorphous superionic materials.[151] There has been a lot of interest in developing superionic polymer materials [152, 153] for battery and other applications currently being employed by their crystalline counterparts. Significant enhancements of conductivity relative to Nernst-Einstein predictions have been observed[154], but the intrinsically slow dynamics of polymeric materials has made it difficult to obtain materials with values comparable to small molecule fluids. Recently, it has been suggested that lowering the dielectric constant of the polymer matrix [155] should greatly improve the situation and other strategies are actively being pursued. One especially promising approach to developing materials of practical interest is being pursued by Sokolov and coworkers who are exploring stiffening the polymer matrix [156] to create a better immobile polymer matrix with a more open pore space in which the mobile ions can move. We suggest this direction, in conjunction with engineering the dielectric properties of the polymer, offers a promising line of attack for creating practical superionic polymer materials for diverse proposed applications. Recent modeling of the dynamics of glass-forming polymer fluids [121, 157] indicates that any molecular factor that changes the packing efficiency of the chains should increase the strength of the temperature dependence of the fluid configurational entropy, amplifying the variation of the scale of collective motion with temperature so that modifying the geometrical structure of the monomers [158] should provide another useful strategy for enhancing superionic conduction in polymer materials. The immobile polymer component must be formed into a stable matrix as in the crystalline superionic material to optimize the capacity for mobile ion transport. We look forward to studying the dynamics of superionic polymer materials in the future within the framework of the string model of relaxation.

**SUPPLEMENTARY MATERIAL**

See supplementary material for Debye Waller factors and Boson peaks.

**ACKNOWLEDGMENTS**

H.Z, and X.Y. W. gratefully acknowledge the support of the Natural Sciences and Engineering Research Council of Canada under the Discovery Grant Program.

**REFERENCE:**

1.      Angell CA. Formation of glasses from liquids and biopolymers. *Science* **267**, 1924-1935 (1995).

2.      Ediger MD, Angell CA, Nagel SR. Supercooled liquids and glasses. *J Phys Chem* **100**, 13200-13212 (1996).

3.      Faraday M. Faraday's diary.  (eds Martin T). G. Bell and sons, ltd (1932).

4.      Chadwick AV. Transport in defective ionic materials: from bulk to nanocrystals. *Phys Status Solidi A* **204**, 631-641 (2007).

5.      Luo W*, et al.* Dynamical stability of body center cubic iron at the Earth's core conditions. *Proc Natl Acad Sci U S A* **107**, 9962-9964 (2010).

6.      Belonoshko AB, Lukinov T, Fu J, Zhao JJ, Davis S, Simak SI. Stabilization of body-centred cubic iron under inner-core conditions. *Nat Geosci* **10**, 312-+ (2017).

7.      Vocadlo L, Alfe D, Gillan MJ, Wood IG, Brodholt JP, Price GD. Possible thermal and chemical stabilization of body-centred-cubic iron in the Earth's core. *Nature* **424**, 536-539 (2003).

8.      Schwegler E, Sharma M, Gygi F, Galli G. Melting of ice under pressure. *Proc Natl Acad Sci U S A* **105**, 14779-14783 (2008).

9.      Millot M*, et al.* Experimental evidence for superionic water ice using shock compression. *Nat Phys* **14**, 297 (2018).

10.     Cavazzoni C, Chiarotti GL, Scandolo S, Tosatti E, Bernasconi M, Parrinello M. Superionic and metallic states of water and ammonia at giant planet conditions. *Science* **283**, 44-46 (1999).

11.     Sun JM, Clark BK, Torquato S, Car R. The phase diagram of high-pressure superionic ice. *Nat Commun* **6**, 8156 (2015).

12.     Matsui M, Price GD. Simulation of the pre-melting behavior of $MgSiO_3$ perovskite at high-pressures and temperatures. *Nature* **351**, 735-737 (1991).

13.     Okeeffe M, Bovin JO. Solid Electrolyte Behavior of $NaMgF_3$ - Geophysical Implications. *Science* **206**, 599-600 (1979).

14.     Oganov AR, Brodholt JP, Price GD. The elastic constants of $MgSiO_3$ perovskite at pressures and temperatures of the Earth's mantle. *Nature* **411**, 934-937 (2001).

15.     Skinner LB, *et al.* Molten uranium dioxide structure and dynamics. *Science* **346**, 984-987 (2014).

16.     Navrotsky A. Taking the measure of molten uranium oxide. *Science* **346**, 916-917 (2014).

17.     Donati C, Douglas JF, Kob W, Plimpton SJ, Poole PH, Glotzer SC. Stringlike cooperative motion in a supercooled liquid. *Phys Rev Lett* **80**, 2338-2341 (1998).

18.     Donati C, Glotzer SC, Poole PH, Kob W, Plimpton SJ. Spatial correlations of mobility and immobility in a glass-forming Lennard-Jones liquid. *Phys Rev E* **60**, 3107-3119 (1999).

19.     Riggleman RA, Yoshimoto K, Douglas JF, de Pablo JJ. Influence of confinement on the fragility of antiplasticized and pure polymer films. *Phys Rev Lett* **97**, 045502 (2006).

20.     Douglas JF, Dudowicz J, Freed KF. Does equilibrium polymerization describe the dynamic heterogeneity of glass-forming liquids? *J Chem Phys* **125**, 144907 (2006).

21.     Zhang H, Kalvapalle P, Douglas JF. String-like collective atomic motion in the melting and freezing of nanoparticles. *J Phys Chem B* **115**, 14068-14076 (2011).

22.     Jin ZH, Gumbsch P, Lu K, Ma E. Melting mechanisms at the limit of superheating. *Phys Rev Lett* **87**, 055703 (2001).

23.     Bai XM, Li M. Ring-diffusion mediated homogeneous melting in the superheating regime. *Phys Rev B* **77**, 134109 (2008).

24.     Zhang H, Khalkhali M, Liu QX, Douglas JF. String-like cooperative motion in homogeneous melting. *J Chem Phys* **138**, 12A538 (2013).

25.     Annamareddy A, Eapen J. Low Dimensional String-like Relaxation Underpins Superionic Conduction in Fluorites and Related Structures. *Sci Rep-Uk* **7**, 44149 (2017).

26.     Annamareddy A, Eapen J. Disordering and dynamic self-organization in stoichiometric $UO_2$ at high temperatures. *J Nucl Mater* **483**, 132-141 (2017).

27.     Annamareddy A, Eapen J. Ion hopping and constrained Li diffusion pathways in the superionic state of antifluorite $Li_2O$. *Entropy-Switz* **19**, 227 (2017).

28.     Zhang H, *et al.* Role of string-like collective atomic motion on diffusion and structural relaxation in glass forming Cu-Zr alloys. *J Chem Phys* **142**, 164506 (2015).

29.     Starr FW, Douglas JF, Sastry S. The relationship of dynamical heterogeneity to the Adam-Gibbs and random first-order transition theories of glass formation. *J Chem Phys* **138**, 12A541 (2013).

30. Wang Y, *et al.* Design principles for solid-state lithium superionic conductors. *Nat Mater* **14**, 1026 (2015).

31. Salanne M, Marrocchelli D, Watson GW. Cooperative mechanism for the diffusion of Li+ ions in LiMgSO$_4$F. *J Phys Chem C* **116**, 18618-18625 (2012).

32. Kamaya N, *et al.* A lithium superionic conductor. *Nat Mater* **10**, 682-686 (2011).

33. Bailey TP, Uher C. Potential for superionic conductors in thermoelectric applications. *Curr Opin Green Sust* **4**, 58-63 (2017).

34. Liu HL, *et al.* Copper ion liquid-like thermoelectrics. *Nat Mater* **11**, 422-425 (2012).

35. Ruello P, Petot-Ervas G, Petot C. Electrical conductivity and thermoelectric power of uranium dioxide. *J Am Ceram Soc* **88**, 604-611 (2005).

36. Gofryk K, *et al.* Anisotropic thermal conductivity in uranium dioxide. *Nat Commun* **5**, 4551 (2014).

37. Qin MJ, *et al.* Thermal conductivity and energetic recoils in UO$_2$ using a many-body potential model. *J Phys-Condens Mat* **26**, 495401 (2014).

38. Foiles SM, Baskes MI, Daw MS. Embedded-Atom-Method functions for the fcc metals Cu, Ag, Au, Ni, Pd, Pt, and their alloys. *Phys Rev B* **33**, 7983-7991 (1986).

39. Parrinello M, Rahman A. Polymorphic transitions in single-crystals - a new molecular-dynamics Method. *J Appl Phys* **52**, 7182-7190 (1981).

40. Nose S. A unified formulation of the constant temperature molecular-dynamics methods. *J Chem Phys* **81**, 511-519 (1984).

41. Hoover WG. Canonical Dynamics - Equilibrium Phase-Space Distributions. *Phys Rev A* **31**, 1695-1697 (1985).

42. Allen MP, Tildesley DJ. *Computer simulation of liquids*. Clarendon Press (1987).

43. Plimpton S. Fast parallel algorithms for short-range molecular-dynamics. *J Comput Phys* **117**, 1-19 (1995).

44. Cooper MWD, Rushton MJD, Grimes RW. A many-body potential approach to modelling the thermomechanical properties of actinide oxides. *J Phys-Condens Mat* **26**, 105401 (2014).

45.     Manara D, Ronchi C, Sheindlin M, Lewis M, Brykin M. Melting of stoichiometric and hyperstoichiometric uranium dioxide. *J Nucl Mater* **342**, 148-163 (2005).

46.     Tammann GHJA. *Lehrbuch der metallkunde; chemie und physik der metalle und ihrer legierungen*, 4. erweiterte aufl., edn. Voss, L. (1932).

47.     Merkle R, Maier J. On the Tannnann-rule. *Z Anorg Allg Chem* **631**, 1163-1166 (2005).

48.     Golunski SE. Why use platinum in catalytic converters? *Platin Met Rev* **51**, 162-162 (2007).

49.     Rapaport DC. *The art of molecular dynamics simulation*, 2nd edn. Cambridge University Press (2004).

50.     Clausen K, Hayes W, Macdonald JE, Osborn R, Hutchings MT. Observation of oxygen Frenkel disorder in uranium-dioxide above 2000-K by use of neutron-scattering techniques. *Phys Rev Lett* **52**, 1238-1241 (1984).

51.     Eapen J, Annamareddy A. Entropic crossovers in superionic fluorites from specific heat. *Ionics* **23**, 1043-1047 (2017).

52.     Goel P, Choudhury N, Chaplot SL. Fast ion diffusion, superionic conductivity and phase transitions of the nuclear materials $UO_2$ and $Li_2O$. *J Phys-Condens Mat* **19**, 386239 (2007).

53.     Rice MJ, Strassler S, Toombs GA. Superionic conductors - theory of phase-transition to cation disordered state. *Phys Rev Lett* **32**, 596-599 (1974).

54.     Parrinello M, Rahman A, Vashishta P. Structural transitions in superionic conductors. *Phys Rev Lett* **50**, 1073-1076 (1983).

55.     Boyce JB, Hayes TM, Mikkelsen JC. Extended-X-ray-absorption-fine-structure investigation of mobile-ion density in superionic AgI, CuI, Cubr, and CuCl. *Phys Rev B* **23**, 2876-2896 (1981).

56.     Annamareddy A, Eapen J. Mobility propagation and dynamic facilitation in superionic conductors. *J Chem Phys* **143**, 194502 (2015).

57.     Cox UJ, Gibaud A, Cowley RA. Effect of impurities on the 1st-order phase-transition of $KMnF_3$. *Phys Rev Lett* **61**, 982-985 (1988).

58.     Ralph J, Hyland GJ. Empirical confirmation of a Bredig transition in $UO_2$. *J Nucl Mater* **132**, 76-79 (1985).

59.    Dworkin AS, Bredig MA. Diffuse transition and melting in fluorite and anti-fluorite yype of compounds - heat content of potassium sulfide from 298 to 1260 degrees K. *J Phys Chem-Us* **72**, 1277 (1968).

60.    Boyce JB, Huberman BA. Superionic conductors - transitions, structures, dynamics. *Phys Rep* **51**, 189-265 (1979).

61.    Wang JC, Gaffari M, Choi S. Ionic-conduction in beta-alumina - potential-energy curves and conduction mechanism. *J Chem Phys* **63**, 772-778 (1975).

62.    Varanasi SR, Yashonath S. Variation of diffusivity with the cation radii in molten salts of superionic conductors containing iodine anion: A molecular dynamics study. *J Chem Sci* **124**, 159-166 (2012).

63.    Voronin BM, Volkov SV. Ionic conductivity of fluorite type crystals $CaF_2$, $SrF_2$, $BaF_2$, and $SrCl_2$ at high temperatures. *J Phys Chem Solids* **62**, 1349-1358 (2001).

64.    Hull S, Norberg ST, Ahmed I, Eriksson SG, Mohn CE. High temperature crystal structures and superionic properties of $SrCl_2$, $SrBr_2$, $BaCl_2$ and $BaBr_2$. *J Solid State Chem* **184**, 2925-2935 (2011).

65.    Sanders D, Eapen J. Thermal jamming of ions in the superionic state of $UO_2$. *Mrs Adv* **3**, 1777-1781 (2018).

66.    Hull S. Superionics: crystal structures and conduction processes. *Rep Prog Phys* **67**, 1233-1314 (2004).

67.    Yakub E, Ronchi C, Staicu D. Molecular dynamics simulation of premelting and melting phase transitions in stoichiometric uranium dioxide. *J Chem Phys* **127**, 094508 (2007).

68.    Lunev AV, Tarasov BA. A classical molecular dynamics study of the correlation between the Bredig transition and thermal conductivity of stoichiometric uranium dioxide. *J Nucl Mater* **415**, 217-221 (2011).

69.    Ye YY, Chen Y, Ho KM, Harmon BN, Lindgard PA. Phonon-phonon coupling and the stability of the high-temperature bcc phase of Zr. *Phys Rev Lett* **58**, 1769-1772 (1987).

70.    Belonoshko AB, Ahuja R, Johansson B. Molecular dynamics study of melting and fcc-bcc transitions in Xe. *Phys Rev Lett* **87**, 165505 (2001).

71.    Belonoshko AB, Ahuja R, Johansson B. Stability of the body-centred-cubic phase of iron in the Earth's inner core. *Nature* **424**, 1032-1034 (2003).

72.     Souvatzis P, Eriksson O, Katsnelson MI, Rudin SP. Entropy driven stabilization of energetically unstable crystal structures explained from first principles theory. *Phys Rev Lett* **100**, 095901 (2008).

73.     Heitjans P, Indris S. Diffusion and ionic conduction in nanocrystalline ceramics. *J Phys-Condens Mat* **15**, R1257-R1289 (2003).

74.     Chadwick AV, Savin SLP. Structure and dynamics in nanoionic materials. *Solid State Ionics* **177**, 3001-3008 (2006).

75.     Beaman RG. Relation between (Apparent) 2nd-Order Transition Temperature and Melting Point. *J Polym Sci* **9**, 470-472 (1952).

76.     Wang LM, Angell CA, Richert R. Fragility and thermodynamics in nonpolymeric glass-forming liquids. *J Chem Phys* **125**, 074505 (2006).

77.     Gillan MJ. Collective dynamics in super-ionic CaF$_2$. 1. simulation compared with neutron-scattering experiment. *J Phys C Solid State* **19**, 3391-3411 (1986).

78.     Gillan MJ. Collective dynamics in super-ionic CaF$_2$. 2. defect interpretation. *J Phys C Solid State* **19**, 3517-3533 (1986).

79.     Gray-Weale A, Madden PA. Dynamical arrest in superionic crystals and supercooled liquids. *J Phys Chem B* **108**, 6624-6633 (2004).

80.     Gray-Weale A, Madden PA. The energy landscape of a fluorite-structured superionic conductor. *J Phys Chem B* **108**, 6634-6642 (2004).

81.     Shiba H, Onuki A, Araki T. Structural and dynamical heterogeneities in two-dimensional melting. *Epl-Europhys Lett* **86**, 66004 (2009).

82.     Delogu F. Homogeneous melting of metals with different crystalline structure. *J Phys-Condens Mat* **18**, 5639-5653 (2006).

83.     Ainslie NG, Mackenzie JD, Turnbull D. Melting kinetics of quartz and cristobalite. *J Phys Chem-Us* **65**, 1718 (1961).

84.     Feng Y, Goree J, Liu U. Solid superheating observed in two-dimensional strongly coupled dusty plasma. *Phys Rev Lett* **100**, 205007 (2008).

85.     Zykova-Timan T, Ceresoli D, Tartaglino U, Tosatti E. Physics of solid and liquid alkali halide surfaces near the melting point. *J Chem Phys* **123**, 164701 (2005).

86.    Kincs J, Martin SW. Non-Arrhenius conductivity in glass: Mobility and conductivity saturation effects. *Phys Rev Lett* **76**, 70-73 (1996).

87.    Okada Y, Ikeda M, Aniya M. Non-Arrhenius ionic conductivity in solid electrolytes: A theoretical model and its relation with the bonding nature. *Solid State Ionics* **281**, 43-48 (2015).

88.    Arakawa S, Shiotsu T, Hayashi S. Non-Arrhenius temperature dependence of conductivity in lanthanum lithium tantalate. *J Ceram Soc Jpn* **113**, 317-319 (2005).

89.    Leon C, Santamaria J, Paris MA, Sanz J, Ibarra J, Varez A. Non-Debye conductivity relaxation in the non-Arrhenius $Li_{0.5}La_{0.5}TiO_3$ fast ionic conductor. A nuclear magnetic resonance and complex impedance study. *J Non-Cryst Solids* **235**, 753-760 (1998).

90.    Ueno K, Zhao ZF, Watanabe M, Angell CA. Protic ionic liquids based on decahydroisoquinoline: lost superfragility and ionicity-fragility correlation. *J Phys Chem B* **116**, 63-70 (2012).

91.    Anouti M, Caillon-Caravanier M, Dridi Y, Galiano H, Lemordant D. Synthesis and characterization of new pyrrolidinium based protic ionic liquids. good and superionic liquids. *J Phys Chem B* **112**, 13335-13343 (2008).

92.    Greaves GN, Meneau F, Majerus O, Jones DG, Taylor J. Identifying vibrations that destabilize crystals and characterize the glassy state. *Science* **308**, 1299-1302 (2005).

93.    Li Y, Bai HY, Wang WH. Low-temperature specific-heat anomalies associated with the boson peak in CuZr-based bulk metallic glasses. *Phys Rev B* **74**, 052201 (2006).

94.    Kaya D, Green NL, Maloney CE, Islam MF. Normal modes and density of states of disordered colloidal solids. *Science* **329**, 656-658 (2010).

95.    Zhang H, Douglas JF. Glassy interfacial dynamics of Ni nanoparticles: Part II Discrete breathers as an explanation of two-level energy fluctuations. *Soft Matter* **9**, 1266-1280 (2013).

96.    Kobayashi M, Tomoyose T, Aniya M. Low energy excitation in cation superionic conductors. *Physica B* **219-20**, 460-462 (1996).

97.    Hoshino S. Structure and dynamics of solid-state ionics. *Solid State Ionics* **48**, 179-201 (1991).

98.    Nakamura M*, et al.* Low energy vibrational excitations characteristic of superionic glass. *Physica B* **385**, 552-554 (2006).

99.    Tse JS*, et al.* Origin of low-frequency local vibrational modes in high density amorphous ice. *Phys Rev Lett* **85**, 3185-3188 (2000).

100. Zhang H, Srolovitz DJ, Douglas JF, Warren JA. Atomic motion during the migration of general 001 tilt grain boundaries in Ni. **55**, 4527-4533 (2007).

101. Yokota I. On deviation from Einstein relation observed for diffusion of Ag+ ions in alpha-$Ag_2S$ and others. *J Phys Soc Jpn* **21**, 420 (1966).

102. Wolf ML. Observation of solitary-wave conduction in a molecular dynamics simulation of the superionic conductor $Li_3N$. *J Phys C Solid State* **17**, L285-L288 (1984).

103. Bishop AR. Non-linear collective excitations in superionic conductors. *J Phys C Solid State* **11**, L329-L335 (1978).

104. ZhengJohansson JXM, McGreevy RL. A molecular dynamics study of ionic conduction in CuI. II. Local ionic motion and conduction mechanisms. *Solid State Ionics* **83**, 35-48 (1996).

105. Xu M, Ding J, Ma E. One-dimensional stringlike cooperative migration of lithium ions in an ultrafast ionic conductor. *Appl Phys Lett* **101**, 031901 (2012).

106. Zhang H, Srolovitz DJ, Douglas JF, Warren JA. Grain boundaries exhibit the dynamics of glass-forming liquids. *P Natl Acad Sci USA* **106**, 7735-7740 (2009).

107. Zhang H, Kalvapalle P, Douglas JF. String-like collective atomic motion in the interfacial dynamics of nanoparticles. *Soft Matter* **6**, 5944-5955 (2010).

108. Betancourt BAP, Douglas JF, Starr FW. String model for the dynamics of glass-forming liquids. *J Chem Phys* **140**, 204509 (2014).

109. Bershtein VA, Egorov VM, Egorova LM, Ryzhov VA. The role of thermal-analysis in revealing the common molecular nature of transitions in polymers. *Thermochim Acta* **238**, 41-73 (1994).

110. Yamamuro O, Tsukushi I, Lindqvist A, Takahara S, Ishikawa M, Matsuo T. Calorimetric study of glassy and liquid toluene and ethylbenzene: Thermodynamic approach to spatial heterogeneity in glass-forming molecular liquids. *J Phys Chem B* **102**, 1605-1609 (1998).

111. Tatsumi S, Aso S, Yamamuro O. Thermodynamic study of simple molecular glasses: universal features in their heat capacity and the size of the cooperatively rearranging regions. *Phys Rev Lett* **109**, 045701 (2012).

112. Betancourt BAP, Hanakata PZ, Starr FW, Douglas JF. Quantitative relations between cooperative motion, emergent elasticity, and free volume in model glass-forming polymer materials. *P Natl Acad Sci USA* **112**, 2966-2971 (2015).

113.   Hanakata PZ, Betancourt BAP, Douglas JF, Starr FW. A unifying framework to quantify the effects of substrate interactions, stiffness, and roughness on the dynamics of thin supported polymer films. *J Chem Phys* **142**, 234907 (2015).

114.   Rivera A, Santamaria J, Leon C, Blochowicz T, Gainaru C, Rossler EA. Temperature dependence of the ionic conductivity in $Li_{3x}La_{2/3-x}TiO_3$: Arrhenius versus non-Arrhenius. *Appl Phys Lett* **82**, 2425-2427 (2003).

115.   Leon C, Santamaria J, Paris MA, Sanz J, Ibarra J, Torres LM. Non-Arrhenius conductivity in the fast ionic conductor $Li_{0.5}La_{0.5}TiO_3$: Reconciling spin-lattice and electrical-conductivity relaxations. *Phys Rev B* **56**, 5302-5305 (1997).

116.   Zhao J, Simon SL, McKenna GB. Using 20-million-year-old amber to test the super-Arrhenius behaviour of glass-forming systems. *Nat Commun*

*Nat Commun* **4**, 1783 (2013).

117.   O'Connell PA, McKenna GB. Arrhenius-type temperature dependence of the segmental relaxation below Tg. *J Chem Phys* **110**, 11054-11060 (1999).

118.   Novikov VN, Sokolov AP. Qualitative change in structural dynamics of some glass-forming systems. *Phys Rev E* **92**, 062304 (2015).

119.   Gupta PK, Heuer A. Physics of the iso-structural viscosity. *J Non-Cryst Solids* **358**, 3551-3558 (2012).

120.   Richert R, Angell CA. Dynamics of glass-forming liquids. V. On the link between molecular dynamics and configurational entropy. *J Chem Phys* **108**, 9016-9026 (1998).

121.   Dudowicz J, Freed KF, Douglas JF. Generalized entropy theory of polymer glass formation. *Adv Chem Phys* **137**, 125-222 (2008).

122.   Sata N, Eberman K, Eberl K, Maier J. Mesoscopic fast ion conduction in nanometre-scale planar heterostructures. *Nature* **408**, 946-949 (2000).

123.   Knoner G, Reimann K, Rower R, Sodervall U, Schaefer HE. Enhanced oxygen diffusivity in interfaces of nanocrystalline $ZrO_2$ center dot $Y_2O_3$. *P Natl Acad Sci USA* **100**, 3870-3873 (2003).

124.   Garcia-Barriocanal J, *et al.* Colossal ionic conductivity at interfaces of epitaxial $ZrO_2$ : $Y_2O_3$/$SrTiO_3$ heterostructures. *Science* **321**, 676-680 (2008).

125.   Liu W, *et al.* Suppressed phase transition and giant ionic conductivity in $La_2Mo_2O_9$ nanowires. *Nat Commun* **6**, 8354 (2015).

126.  Zhang H, Yang Y, Douglas JF. Influence of string-like cooperative atomic motion on surface diffusion in the (110) interfacial region of crystalline Ni. *J Chem Phys* **142**, 084704 (2015).

127.  Wang XY, Tong XH, Zhang H, Douglas JF. String-like collective motion and diffusion in the interfacial region of ice. *J Chem Phys* **147**, 194508 (2017).

128.  Rhead GE. On surface diffusion and existence of 2-dimensional liquids. *Surf Sci* **15**, 353-357 (1969).

129.  Binh VT, Melinon P. On viscous mechanism for surface-diffusion at high-tempeartures (T/Tm greater than 0.75) due to formation of a 2d dense fluid on metallic surdaces. *Surf Sci* **161**, 234-244 (1985).

130.  Bonzel HP, Latta EE. Surface self-diffusion on Ni(110) - temperature-dependence and directional anisotropy. *Surf Sci* **76**, 275-295 (1978).

131.  Truhlar DG, Kohen A. Convex Arrhenius plots and their interpretation. *P Natl Acad Sci USA* **98**, 848-851 (2001).

132.  Nagel ZD, Dong M, Bahnson BJ, Klinman JP. Impaired protein conformational landscapes as revealed in anomalous Arrhenius prefactors. *P Natl Acad Sci USA* **108**, 10520-10525 (2011).

133.  Liang ZX, Lee T, Resing KA, Ahn NG, Klinman JP. Thermal-activated protein mobility and its correlation with catalysis in thermophilic alcohol dehydrogenase. *P Natl Acad Sci USA* **101**, 9556-9561 (2004).

134.  Zavodszky P, Kardos J, Svingor A, Petsko GA. Adjustment of conformational flexibility is a key event in the thermal adaptation of proteins. *P Natl Acad Sci USA* **95**, 7406-7411 (1998).

135.  Haddadian EJ, Zhang H, Freed KF, Douglas JF. Comparative study of the collective dynamics of proteins and inorganic nanoparticles. *Sci Rep-Uk* **7**, 41671 (2017).

136.  Castiglione MJ, Madden PA. Fluoride ion disorder and clustering in superionic $PbF_2$. *J Phys-Condens Mat* **13**, 9963-9983 (2001).

137.  Haven Y. Concentration and association of lattice defects in NaCl. In: *Report of the conference on defects in crystalline solids* (ed^(eds). Physical Society (1954).

138.  Annamareddy VA, Nandi PK, Mei XJ, Eapen J. Waxing and waning of dynamical heterogeneity in the superionic state. *Phys Rev E* **89**, 010301(R) (2014).

139.    Freed KF. Communication: Towards first principles theory of relaxation in supercooled liquids formulated in terms of cooperative motion. *J Chem Phys* **141**, 141102 (2014).

140.    Sciortino F, Geiger A, Stanley HE. Effect of defects on molecular mobility in liquid water. *Nature* **354**, 218-221 (1991).

141.    Granato AV. Thermodynamic and kinetic properties of amorphous and liquid states. *Metall Mater Trans A* **29**, 1837-1843 (1998).

142.    Norby T. The promise of protonics. *Nature* **410**, 877-878 (2001).

143.    Hayashi A, Noi K, Sakuda A, Tatsumisago M. Superionic glass-ceramic electrolytes for room-temperature rechargeable sodium batteries. *Nat Commun* **3**, 856 (2012).

144.    Bouchet R, *et al.* Single-ion BAB triblock copolymers as highly efficient electrolytes for lithium-metal batteries. *Nat Mater* **12**, 452-457 (2013).

145.    Miller TA, Wittenberg JS, Wen H, Connor S, Cui Y, Lindenberg AM. The mechanism of ultrafast structural switching in superionic copper (I) sulphide nanocrystals. *Nat Commun* **4**, 1369 (2013).

146.    Chen P, Xiong ZT, Luo JZ, Lin JY, Tan KL. Interaction of hydrogen with metal nitrides and imides. *Nature* **420**, 302-304 (2002).

147.    Kang B, Ceder G. Battery materials for ultrafast charging and discharging. *Nature* **458**, 190-193 (2009).

148.    Maekawa H, *et al.* Halide-stabilized $LiBH_4$, a room-temperature lithium fast-ion conductor. *J Am Chem Soc* **131**, 894 (2009).

149.    Mizuno F, Belieres JP, Kuwata N, Pradel A, Ribes M, Angell CA. Highly decoupled ionic and protonic solid electrolyte systems, in relation to other relaxing systems and their energy landscapes. *J Non-Cryst Solids* **352**, 5147-5155 (2006).

150.    Ricco M, *et al.* Superionic conductivity in the $Li_4C_{60}$ fulleride polymer. *Phys Rev Lett* **102**, 145901 (2009).

151.    Minami T. Recent progress in superionic conducting glasses. *J Non-Cryst Solids* **95-6**, 107-118 (1987).

152.    Lin KJ, Maranas JK. Superionic behavior in polyethylene-oxide-based single-ion conductors. *Phys Rev E* **88**, 052602 (2013).

153.    Wang YY, *et al.* Examination of the fundamental relation between ionic transport and segmental relaxation in polymer electrolytes. *Polymer* **55**, 4067-4076 (2014).

154.    LaFemina NH, Chen Q, Colby RH, Mueller KT. The diffusion and conduction of lithium in poly(ethylene oxide)-based sulfonate ionomers. *J Chem Phys* **145**, 114903 (2016).

155.    Wheatle BK, Keith JR, Mogurampelly S, Lynd NA, Ganesan V. Influence of dielectric constant on ionic transport in polyether-based electrolytes. *Acs Macro Lett* **6**, 1362-1367 (2017).

156.    Wojnarowska Z, *et al.* Effect of chain rigidity on the decoupling of ion motion from segmental relaxation in polymerized ionic liquids: ambient and elevated pressure studies. *Macromolecules* **50**, 6710-6721 (2017).

157.    Stukalin EB, Douglas JF, Freed KF. Application of the entropy theory of glass formation to poly(alpha-olefins). *J Chem Phys* **131**, 114905 (2009).

158.    Fan F, Wang YY, Hong T, Heres MF, Saito T, Sokolov AP. Ion conduction in polymerized ionic liquids with different pendant groups. *Macromolecules* **48**, 4461-4470 (2015).